\documentclass[12pt,preprint]{aastex}
\usepackage{amsmath}
\usepackage{color}
\newcommand{\del}[2]%
{\frac{\mathrm{d}{#2}}{\mathrm{d}{#1}}}
\newcommand{\Del}[2]%
{\frac{\mathrm{D}{#2}}{\mathrm{D}{#1}}}
\newcommand{\ddel}[2]%
{\frac{\mathrm{d}^2{#2}}{\mathrm{d}{#1}^2}}
\newcommand{\pdel}[2]%
{\frac{\partial{#2}}{\partial{#1}}}
\newcommand{\pddel}[2]%
{\frac{\partial^2{#2}}{\partial{#1}^2}}

\newcommand{\Ms}{M_{\odot}}
\newcommand{\km}{\,\, \mathrm{km}}

\newcommand{\cmps}{\,\, \mathrm{cm \,\, s^{-1}}}

\newcommand{\gauss}{\,\, \mathrm{G}}

\newcommand{\psec}{\,\, \mathrm{s^{-1}}}
\newcommand{\gpcmc}{\,\, \mathrm{g \,\, cm^{-3}}}

\newcommand{\radps}{\,\, \mathrm{rad \,\, s^{-1}}}

\newcommand{\ergps}{\,\, \rm erg \,\, {s}^{-1}}

\slugcomment{Accepted to the ApJ}
\shorttitle{Neutrino Pair Annihilation in Collapsars}
\shortauthors{Harikae et al.}

\begin{document}

\title{Neutrino Pair Annihilation in Collapsars; \\
 Ray-Tracing Method in Special Relativity}

\author{Seiji Harikae\altaffilmark{1,2},
 Kei Kotake \altaffilmark{2,3}, 
and Tomoya Takiwaki\altaffilmark{2}}

\affil{\altaffilmark{1}Department of Astronomy, The Graduate School of Science,
University of Tokyo, 7-3-1 Hongo, Bunkyo-ku,Tokyo, 113-0033, Japan}
\affil{\altaffilmark{2}Division of Theoretical Astronomy, 
  National Astronomical Observatory of Japan, 2-21-1, 
  Osawa, Mitaka, Tokyo, 181-8588, Japan}
\affil{\altaffilmark{3}Center for Computational Astrophysics,
 National Astronomical Observatory of Japan, 2-21-1, 
  Osawa, Mitaka, Tokyo, 181-8588, Japan}

\email{seiji.harikae@nao.ac.jp}

\begin{abstract}
 We develop a numerical scheme and 
code for estimating the energy and momentum transfer via 
 neutrino pair annihilation ($\nu + {\bar \nu} \rightarrow e^{-}+ e^{+}$),
 bearing in mind the application to the collapsar models of gamma-ray bursts (GRBs).
 To calculate the neutrino flux illuminated from the accretion disk, 
we perform a ray-tracing calculation in the framework of 
special relativity. 
The numerical accuracy of the developed code is certificated by several
 tests, in which we show comparisons with the corresponding analytical solutions. 
Using hydrodynamical data in our 
collapsar simulation, we estimate the annihilation rates in a post-processing manner.
 We show that the neutrino energy deposition and momentum transfers are 
 strongest near the inner edge of the accretion disk. The beaming 
effects of special relativity are found to change the annihilation 
rates by several factors in the polar funnel region.
 After the accretion disk settles into a stationary state (typically later than 
 $\sim 9$ s from the onset of gravitational collapse),
 we find that 
the neutrino-heating timescale in the vicinity of the polar funnel ($\lesssim 80$ km)
can become shorter than the hydrodynamical timescale, indicating
 that the neutrino-heated outflows can be launched there. We point out that the 
momentum transfer can play as important role as the energy deposition for the 
 efficient acceleration of neutrino-driven outflows. 
 Our results suggest that the neutrino pair annihilation 
 has a potential importance equal to the conventional magnetohydrodynamic mechanism
  for igniting the GRB fireballs.
\end{abstract}


\section{Introduction}
 Accumulating observational evidences have been reported so far, identifying 
 a massive stellar collapse as the origins of long-duration gamma-ray bursts (GRBs), 
such as the preponderance of short-lived massive star formation for their host
 galaxies, as well as the identification of SN Ib/c light curves in the afterglows
 \citep{pacz98,gala98,hjor03,stan03}.
The duration of the long bursts may correspond to the accretion of debris falling 
into the central black
hole (BH)\citep{piro98}.
 Pushed by those observations, the so-called collapsar  
 has received quite some interest for their central engines \citep{woos93,pacz98,macf99}.

 In the collapsar model, the central cores 
with significant angular momentum collapse into a BH. 
Neutrinos emitted from the accretion disk heat the matter of the polar funnel region to
launch the GRB outflows. \citet{paz90,mezree} pioneerlingly 
proposed that the energy deposition 
proceeds predominantly
 via neutrino and antineutrino annihilation into electron and positron (e.g., $\nu + {\bar \nu} \rightarrow e^{-}+ e^{+}$, hereafter 
``neutrino pair annihilation''). 
The relativistic outflows are expected to ultimately 
form a fireball, which is 
good for explaining the observed afterglow (e.g.,  \citet{pira99}).
In addition, it is suggested that the strong magnetic fields in the
cores of order of $10^{15} \gauss$ play also an active role both for driving 
the magneto-driven jets and for extracting a significant amount of
energy from the central engine (e.g.,
\cite{whee00,thom04,uzde07a} and see references therein). 

However, it is still controversial whether the generation of the relativistic outflows 
proceeds predominantly via magnetohydrodynamic (MHD) or neutrino-heating processes. 
So far, much attention has been paid to the MHD processes 
(e.g., \citet{prog03b,prog03c,mizu06,fuji06,fuji07,naga07,naga09,hari09}).
A general outcome of these extensive MHD simulations is that the magneto-driven 
shock waves, produced mainly by the field-wrapping of the strong precollapse 
magnetic fields or by the field amplification due to magnetorotational instability 
(MRI), can blow up massive stars along the rotational axis. Although those primary 
jet-like explosions are at most mildly relativistic due to 
too much baryons, they may be related to the X-ray flashes, 
which is a low energy analogue of the GRBs \citep{sode06,ghis07},
 and those jets, which make a low density
 funnel along the rotational axis, could provide a good birthplace for the subsequent 
relativistic outflows \citep{hari09}. 

In contrast to such blossoms in MHD studies, 
 there are only a few studies pursuing the possibility of 
generating jets by the energy deposition via neutrino pair annihilation.
This is mainly because the neutrino emission from 
the accretion disk generally becomes highly aspherical, thus demanding one 
in principle, to solve the multidimensional neutrino transfer problem.
For the first time in the numerical simulations, 
\citet{macf99} pointed out the importance of the energy 
deposition via neutrino pair annihilation 
in their simulations, however the energy deposition 
 rates to the polar funnel region were adjusted by hand to produce jets.
   Without those artificial energy injections, neutrino-driven outflows 
have been yet to be realized so far in the numerical simulations, to our best knowledge.
It should be noted in the MHD-driven collapsars 
that too much precollapse magnetic fields with rapid 
 rotation can lead to the MHD-driven explosion, which can prohibit the formation of a BH
 (e.g., \citet{dess08,bucc09,burr07}), thus not good for the collapsar scenario.
 Therefore it is still important to address the possibility of the neutrino-driven mechanism in collapsars.

Thus far, 
several methods for the implementation of the neutrino pair annihilation, 
have been reported.
One straightforward method is to solve the full radiation field by the Monte-Carlo 
  method \citep{tubbs78,janka89}. Still at present, this seems too expensive 
 to combine with the hydrodynamic simulations.
 By estimating the local fluxes and spectra of the neutrino emission
 via the so-called neutrino leakage scheme,
 \citet{ruff97} proposed to estimate the heating rate by summing up 
 the local contributions of the neutrino and antineutrino radiation incident from 
all directions. 
However this scheme requires the double angular integration for neutrino and antineutrino 
 for all the grid points in three dimension, which is still highly 
 time-consuming and prevents its application to hydrodynamical simulation. 
 \citet{ruff98} have upgraded this scheme to be more feasible, by defining 
 the position of the neutrinosphere. Instead of the three-dimensional stellar volume
 as neutrino and antineutrino sources like in \citet{ruff97}, the energy deposition 
rates can be estimated by summing only over the two-dimensional neutrinospheres, which 
 reduces the computational cost significantly.

Along this prescription, \citet{naga07} have estimated the 
 neutrino heating rates, and included them to the hydrodynamical simulation.
  For reducing the computational time, 
 they added one more assumption of the optically thinness 
of the accretion disk to the prescription by \citet{ruff98}.
Even with this potential 
overestimation of the heating rates, the neutrino-driven outflows
 were not observed in their simulations.
 As for the methodology of \citet{ruff98}, the neutrino fluxes are estimated 
 by summing up the local rates only along the $z-$direction (:perpendicular 
 to the equatorial plane), which simplifies the directional dependence of the incoming
 neutrinos. The directional dependence 
 can be more appropriately handled by doing  
the so-called ray-tracing analysis. 
Moreover, even when the spacial structure of the 
 accretion disk becomes highly inhomogeneous, which is the case for our collapsar 
 simulations, the ray-tracing analysis is good for 
 omitting the surface of the accretion disk, from which the neutrino 
 rays cannot travel, 
such as from the regions opposite to the central black hole.

More recently \citet{dess09} have developed a new scheme 
to estimate the energy deposition rates via neutrino pair annihilation 
using the mutliangle neutrino-transport solver \citep{ott_angle}. They discussed
 the possibility of the neutrino-driven outflow production in the postmerger phase of 
binary neutron-star coalescence. 
Without the general relativistic effects, this is the state-of-the-art method
to estimate the neutrino pair annihilation.
As for the general relativistic effects, \citet{birkl} have performed a series of 
ray-tracing calculations in a Kerr spacetime. They found that the total
 energy deposition rates measured by an observer at infinity can
 increase by a factor of two by the general relativistic effect, which is 
consistent with the result of previous studies (e.g., \citet{asano1,asano2}). 

 Joining in the extensive efforts mentioned above, 
here we present a numerical scheme and code for calculating 
the neutrino heating via pair 
neutrino annihilation, which is designed to be incorporable to the hydrodynamical 
simulations of collapsars.
 Relying on the neutrino leakage scheme as in \citet{ruff97,ruff98}, we estimate 
the incident neutrino flux by doing a ray-tracing calculation from 
the neutrino sphere, on which the neutrino distribution function is 
assumed to take a Fermi-Dirac distribution. 
The ray-tracing calculation is done in the Minkowskian spacetime
 (e.g., \citet{kotake_09}), neglecting 
 the general relativistic (GR) effects 
(e.g., \citet{birkl}) for simplicity.
 Thus our scheme is limited to capture the special relativistic effects 
such as beaming effects. 
One of the main characteristics of our method is that the 
neutrino pair annihilation can be estimated only with quantities 
 defined on the numerical grids of the hydrodynamical simulation. This can be done
 by carrying out a Lorentz transformation of radiation variables in the 
 comoving frame back to the Eulerian frame. By doing so, the computational 
 cost of the double angular integration that is required in estimating the annihilation 
rates, can be reduced significantly.
 We derive the formulation required for this, and 
 describe it elaborately.
 We check the numerical accuracy of the developed code by showing
 the comparisons with analytical solutions, some of 
which we newly give in this paper. 
 Based on the results of our long-term collapsar simulation \citep{hari09},
 we run our new code to estimate the heating rate in a post-processing manner.
 Although the presented scheme is not the state-of-the-art in comparison
 with \citet{birkl,dess09}, we elaborately study the possibility of the 
neutrino-driven mechanism in collapsars and will point out that the neutrino pair 
annihilation has a potential importance equal to the conventional MHD mechanism
  for igniting the GRB fireballs.

This paper is organized as follows. 
We summarize the special relativistic formulation of 
the neutrino energy deposition and momentum transfer in Section \ref{sec:formulation}.
 Section 3 is devoted to the main results. The numerical tests and its comparison 
 with the analytical solutions are shown in section 3.1.
In section \ref{sec:application}, we estimate the annihilation rates in a 
post-processing manner using hydrodynamical data in our 
collapsar simulation.
 We summarize our results and discuss their implications in  Section \ref{sec:discussion}.

\begin{figure}[tbd]
\epsscale{0.8}\plotone{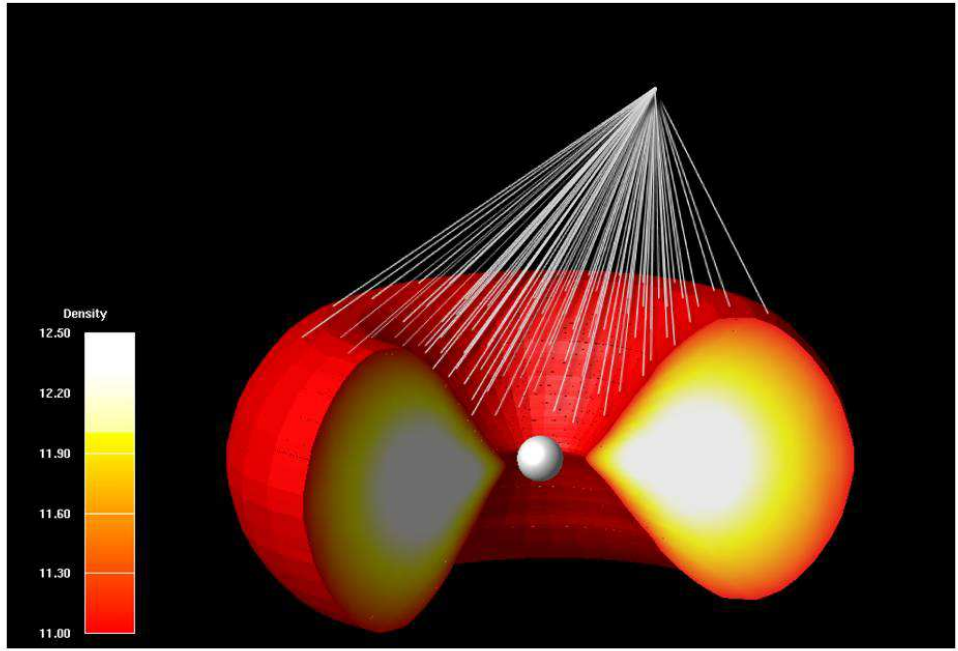}
\caption{An example of the ray-tracing (white line) of neutrinos for estimating the
 neutrino pair annihilation towards a given point outside the accretion disk 
(indicated by red). The central sphere represents the black hole (BH). 
 We trace each neutrino trajectory backwards till 
 it hits to the surface of the neutrino sphere, which closely coincides with the 
 surface of the accretion disk (footpoints of the rays on the torus). 
Note again that the geodesic of neutrinos is given by the 
straight line, since the general relativistic effects are not included in this study.}
\label{fig:3d_heat}
\end{figure} 

\section{Neutrino pair annihilation in special relativity}
\label{sec:formulation}

Before going into details, we first show an illustrative figure, which presents
 our calculation concept for the neutrino pair annihilation in collapsars
 (Figure \ref{fig:3d_heat}). For estimating the
 neutrino pair annihilation at a given target, we trace each neutrino trajectory 
(white lines) backwards till it hits to the surface of the accretion disk (colored by 
red). 
It should be noted that the surface 
 closely coincides with the surface of the neutrino sphere (footpoints of the rays 
on the tori), from which neutrinos and antineutrinos emerge freely out.
  As indicated in this figure, some fraction of neutrinos comes out from the 
 the inner edge of the disk, whose rotational speed is close to the speed of light. 
 This suggests that the 
 special relativistic effects become important towards the precise determination 
of the heating rates. 

\begin{figure}[tbd]
\epsscale{0.6}\plotone{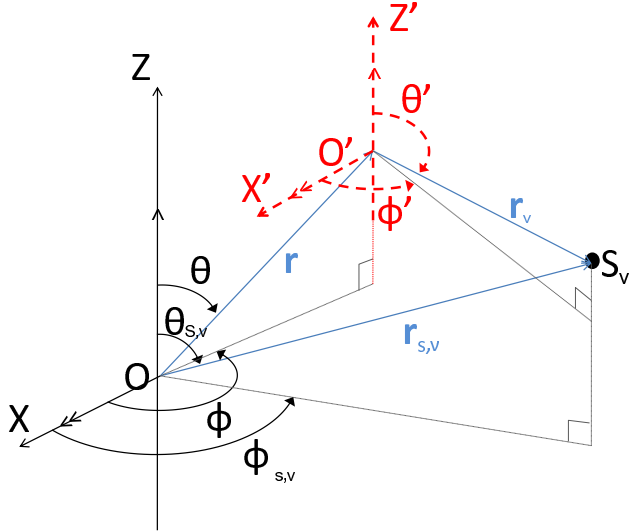}
\caption{Schematic picture of the two coordinate system bridging between 
one for the hydrodynamic quantities (${\bold O}$)
 and another for the radiation quantities (${\bold O}'$).
Note that the coordinate ${\bold O}'$ is used to describe the 
 target position, where the neutrino pair annihilation occurs.
$S_{\nu}$ denotes the neutrino/antineutrino source. 
${\mbox{\boldmath$r$}}_{s,\nu} (r_{s,\nu}, \theta_{s,\nu}, \phi_{s,\nu})$ is the position of $S_{\nu}$ 
measured in ${\bold O}$, while 
${\mbox{\boldmath$r$}}_{\nu} (r_{\nu}, \theta_{\nu}, \phi_{\nu})$ is the position of $S_{\nu}$ 
measured in ${\bold O}'$. 
$\mbox{\boldmath$r$} (r,\theta,\phi)$ is the position of ${\bold O}'$ measured in ${\bold O}$. }
\label{fig:3d_axis}
\end{figure}

For later convenience, we define the two frames namely of ${\bold O}$ and ${\bold O}'$ 
 in the spherical coordinate systems (Figure \ref{fig:3d_axis}). 
Here ${\bold O}$ is the coordinate system that we use for
 describing the variables obtained 
in the hydrodynamical calculations. On the other hand, ${\bold O}'$ is the one for 
describing the target points where the pair annihilation occurs.
 Note that the coordinates between ${\bold O}$ and ${\bold O}'$ system, 
 differ only in the position of their basis. $S_{\nu}$ in the figure represents the position of 
neutrino/antineutrino source. 
 $S_{\nu}$ is usually described by the ${\bold O}'$ system. 
In the following, variables measured in the ${\bold O}'$ 
 system, are described with the subscript $\nu$ or $\bar{\nu}$, indicating that they
 are related to the neutrino sources.
 With these definitions, we move on to describe the special relativistic formulation
  from the next section.

\subsection{Formulation of energy deposition and momentum transfer rates}
\label{subsec:srheat}

The energy deposition rate via neutrino pair annihilation (e.g., \citet{good87,asano2}) 
is written as,
\begin{eqnarray}
    \frac{dq_{\nu {\bar{\nu}}}^+(\mbox{\boldmath$r$})}{dt dV} 
	&=& \iint f_{\nu}(p_{\nu},\mbox{\boldmath$r$}) f_{\bar \nu}(p_{\bar \nu},
\mbox{\boldmath$r$}) 
	\sigma |\mbox{\boldmath$v$}_{\nu} - \mbox{\boldmath$v$}_{\bar \nu}| 
  (\epsilon_{\nu} + \epsilon_{\bar \nu})    
 d^3 \mbox{\boldmath$p$}_{\nu} d^3 \mbox{\boldmath$p$}_{\bar \nu}, \label{eq:heatdef}
\end{eqnarray}
where $f_\nu$ is the number density of neutrinos in phase space, 
$\mbox{\boldmath$p$}_\nu$, $\epsilon_\nu$, $\mbox{\boldmath$v$}_\nu$, 
 is the momentum,  energy, and velocity of neutrino, respectively. 
Those definitions are the same for antineutrino by changing the notation 
$\nu$ to ${\bar \nu}$. $\sigma$ is the cross section given by 
\begin{equation}
\sigma = 2c^2KG_{\rm F}^2(\mbox{\boldmath$p$}_\nu \cdot \mbox{\boldmath$p$}_{\bar \nu}), 
\end{equation}
where the dimensionless parameter $K$ is written as 
\begin{eqnarray}
K(\nu_e, {\bar \nu}_e) = \frac{1+4\sin^2\theta_{\rm W}+8\sin^4\theta_{\rm W}}{6 \pi}, \\ 
K(\nu_\mu, {\bar \nu}_\mu) = K(\nu_\tau, {\bar \nu}_\tau) = \frac{1-4\sin^2\theta_{\rm W}+8\sin^4\theta_{\rm W}}{6 \pi}.  
\end{eqnarray}
Here, the Fermi constant $G_{\rm F}^2=5.29\times10^{-44} {\rm cm^2~MeV^{-2}}$, 
and the Weinberg angle is $\sin^2\theta_{\rm W} = 0.23$. 
Note in our setting of the 
 coordinate system (e.g., Figure 2) that \mbox{\boldmath$r$} corresponds to the
 origin of the ${\bold O}'$ (targets) system measured in the ${\bold O}$ system. 
Variables with subscripts of $\nu$ ($\bar{\nu}$) are quantities upon 
the neutrinosphere, which is labeled by $S_{\nu (\bar{\nu})}$ 
in Figure 2. They are also measured in the ${\bold O}'$ system. 

Aiming at a straightforward incorporation of the ray-tracing calculation 
into the hydrodynamical simulations, 
 we calculate the energy deposition(momentum transfer) rates
 in the laboratory (lab) frame. On the other hand, 
 the quantities related to radiation, such as the distribution function and 
emissivity/absorptivity 
are locally defined in the rest frame of fluid, in which the radiation 
 isotropy is maintained.  Designating the subscript 0 to the variables in the rest 
frame, the Lorentz transformations between the two frames are given in \citet{mm84} as, 
\begin{eqnarray}
    dtdV &=& dt_0dV_0 \label{lor_start} \\
    d\epsilon &=& \frac{\epsilon}{\epsilon_0}d\epsilon_0 \\
    d\Omega &=& \frac{\epsilon_0^2}{\epsilon^2} d\Omega_0 \\
    \epsilon &=& \gamma \left(1+\beta\mu_0\right) \epsilon_0 \nonumber \\
    &=& \epsilon_0 / [\gamma \left(1-\beta\mu\right)],  \label{lor_end}
\end{eqnarray}
where $\beta = |\mbox{\boldmath$V$}|/c, \gamma = (1-\beta^2)^{-1/2}$, and $\mu = \mbox{\boldmath$V$}\cdot\mbox{\boldmath$n$}/|\mbox{\boldmath$V$}|$ 
 with $\mbox{\boldmath$V$}$ and $\mbox{\boldmath$n$}$, being  
the velocity of fluid and the unit vector describing a given ray of neutrino, within a 
solid angle of $d\Omega$.

 Substituting from Equations (\ref{lor_start}) to (\ref{lor_end}) in Equation (1) gives
 the heating rate in the lab frame as, 
\begin{eqnarray}
    \frac{dq_{\nu {\bar{\nu}}}^+(\mbox{\boldmath$r$})}{dt dV}
	&=& 2 c K G_{\rm F}^2 
    \int
    d\theta_{\nu} d \phi_{\nu}  d\theta_{\bar{\nu}} d \phi_{\bar{\nu}} \nonumber \\
    && \times [ \xi_\nu^5 (\mbox{\boldmath$r$},\Omega_\nu) \xi_{\bar{\nu}}^4 (\mbox{\boldmath$r$},\Omega_{\bar{\nu}}) E_{\nu,0}(\mbox{\boldmath$r$},\Omega_\nu) N_{\bar{\nu},0}(\mbox{\boldmath$r$},\Omega_{\bar{\nu}}) \nonumber \\
         && \quad + \xi_\nu^4 (\mbox{\boldmath$r$},\Omega_\nu) \xi_{\bar{\nu}}^5 (\mbox{\boldmath$r$},\Omega_{\bar{\nu}}) N_{\nu,0}(\mbox{\boldmath$r$},\Omega_\nu) E_{\bar{\nu},0}(\mbox{\boldmath$r$},\Omega_{\bar{\nu}})  ]
    \nonumber \\
    && \times \left[ 1-\sin{\theta_{\nu}} \sin{\theta_{\bar{\nu}}}
        \cos{(\varphi_{\nu}-\varphi_{\bar{\nu}})}
        -\cos{\theta_{\nu}} \cos{\theta_{\bar{\nu}}}
    \right]^2
    \sin \theta_{\nu}\sin \theta_{\bar{\nu}}. \label{eq:heat_ann}
\end{eqnarray}
Here $\xi_\nu$ reflects the special relativistic correction to the 
 neutrino energy as 
\begin{eqnarray}
	\xi_\nu(\mbox{\boldmath$r$},\Omega_\nu) &=& \epsilon_\nu / \epsilon_{\nu,0} \label{eq:heatdef_srstart} \nonumber \\
        &=& 1/[\gamma_\nu(1-\mu_{\nu}\beta_{\nu})]. 
\end{eqnarray}
 Here it should be noted again that the variables with the subscript 0 and $\nu$ denote
 the ones defined in the rest frame of the neutrino source and on the neutrino sphere
 respectively, such that 
\begin{equation}
\beta_\nu   = \frac{|\mbox{\boldmath$V$}_\nu|}{c},
\end{equation}
 is the 
 velocity of the fluid on the neutrino sphere with its Lorentz factor of 
\begin{equation}
\gamma_\nu  = \frac{1}{\sqrt{1-\beta_{\nu}^2}},
\end{equation}
 and 
the directional cosine of 
\begin{equation}
\mu_{\nu}   = \frac{\mbox{\boldmath$n$}_\nu\cdot\mbox{\boldmath$V$}_\nu}{|\mbox{\boldmath$V$}_\nu|},
\label{eq:heatdef_srend} 
\end{equation}
where $\mbox{\boldmath$n$}_\nu$ is the direction of the rays of neutrino from the 
neutrino sphere, 
\begin{eqnarray}
	\mbox{\boldmath$n$}_\nu        &=& \frac{\mbox{\boldmath$p$}_\nu}{|\mbox{\boldmath$p$}_\nu|} 
        = (\sin \theta_\nu \cos \phi_\nu,\  
        \sin \theta_\nu \sin \phi_\nu, \ 
        \cos \theta_\nu).
\end{eqnarray}
The following two quantities are the energy-weighted integration of the 
 neutrino distribution function on the neutrino sphere 
(namely $f_{\nu,0}(\mbox{\boldmath$r$}_{\nu, 0}, 
\mbox{\boldmath$p$}_{\nu,0}$)) as,  
\begin{eqnarray}
	E_{\nu,0}(\mbox{\boldmath$r$},\Omega_\nu)  &=& \int \epsilon_{\nu, 0}^4 f_{\nu,0}(\mbox{\boldmath$r$}_{\nu, 0}, \mbox{\boldmath$p$}_{\nu,0}) d\epsilon_{\nu,0}, \label{eq:heatdef2} \\	
	N_{\nu,0}(\mbox{\boldmath$r$},\Omega_\nu)  &=& \int \epsilon_{\nu, 0}^3 f_{\nu,0}(\mbox{\boldmath$r$}_{\nu, 0}, \mbox{\boldmath$p$}_{\nu,0}) d\epsilon_{\nu, 0}, \label{eq:heatdef3}.
\end{eqnarray}
As in \citet{ruff97,ruff98,naga07}, the neutrino number flux is simply assumed to 
 be conserved as $f_{\nu}(\mbox{\boldmath$r$}, \mbox{\boldmath$p$}_\nu)
=  f_\nu(\mbox{\boldmath$r$}_{\nu, 0}, \mbox{\boldmath$p$}_{\nu, 0})$ 
along its trajectory to the target.
The neutrino distribution function on the neutrino sphere 
 is assumed to take a Fermi-Dirac shape with a vanishing chemical potential as,  
\begin{eqnarray}
    f_\nu(\mbox{\boldmath$r$}_{\nu, 0}, \mbox{\boldmath$p$}_{\nu, 0})
    &=& \frac{1}{(hc)^3} \frac{dn_0}{d\epsilon_0 d\Omega_0 dt_0 dV_0} \nonumber \\
	&=&\frac{1}{(hc)^3}\frac{1}{\exp(\epsilon_{\nu,0} / kT_{\nu,0}) + 1}, 
\end{eqnarray}
where $T_{\nu,0}$ is taken to be the same with the matter temperature 
 on the neutrino sphere of $T(\mbox{\boldmath$r$}_\nu)$. 
 Now the energy integrals in Equations (\ref{eq:heatdef2}) and (\ref{eq:heatdef3}) can be 
expressed by the Fermi integrals 
${\cal F}_k$ as 
\begin{eqnarray}
    {\cal F}_k(y) &\equiv& 
    \int \limits_0^\infty
    \frac{x^k}{\exp(x-y) + 1}
    dx, \label{Fk_start} \\
    E_{\nu,0}(\mbox{\boldmath$r$},\Omega) &=& \frac{(kT(\mbox{\boldmath$r$}_\nu))^5}{(hc)^3}{\cal F}_4(0), \\
    N_{\nu,0}(\mbox{\boldmath$r$},\Omega) &=& \frac{(kT(\mbox{\boldmath$r$}_\nu))^4}{(hc)^3}{\cal F}_3(0).  \label{Fk_end}
\end{eqnarray}
The surface of the neutrino spheres is defined 
 per each lateral angular grid in the hydrodynamic calculation ($\theta$) as
\begin{equation}
\int_{R_{\nu_{i}}(\theta)}^{\infty} \frac{1}{\lambda_{i}}~dr = 2/3,
\label{mfp}
\end{equation}
 where $R_{\nu_{i}}(\theta)$ is the radius of the neutrinosphere for the neutrino 
species of $i$, namely for $\nu_e,\bar{\nu}_e$, and $\nu_{\mu,\tau}(\bar{\nu}_{\mu,\nu})$ 
with the corresponding mean free path calculated by a multi-flavor leakage scheme 
(\citet{epst81,ross03,kota03a,taki09}) with special relativistic corrections 
\citep{hari09}.  
As for the neutrino opacities, electron capture on proton and free 
nuclei, positron capture on neutron, neutrino scattering with nucleon and nuclei, 
photo-pair, plasma processes are included (e.g., \citet{ross03}).

Following the same procedure above, the momentum transfer rate in special relativity
 can be also derived as,
\begin{eqnarray}
    \frac{d {\mbox{\boldmath$m$}}_{\nu \bar{\nu}}^+(\mbox{\boldmath$r$})}{dt dV}&=&
	2 K G_{\rm F}^2 
    \int
    d\theta_{\nu} d \phi_{\nu}  d\theta_{\bar{\nu}} d \phi_{\bar{\nu}} \nonumber \\
    && \times  [ \xi_\nu^5 (\mbox{\boldmath$r$},\Omega_\nu) \xi_{\bar{\nu}}^4 (\mbox{\boldmath$r$},\Omega_{\bar{\nu}}) E_{\nu,0}(\mbox{\boldmath$r$},\Omega_\nu) N_{\bar{\nu},0}(\mbox{\boldmath$r$},\Omega_{\bar{\nu}}) 
 \mbox{\boldmath$n$}_\nu  \nonumber \\
	&& \ + \xi_{\bar{\nu}}^5 (\mbox{\boldmath$r$},\Omega_{\bar{\nu}}) \xi_{\nu}^4 (\mbox{\boldmath$r$},\Omega_{\nu}) E_{\bar{\nu},0}(\mbox{\boldmath$r$},\Omega_{\bar{\nu}}) N_{\nu,0}(\mbox{\boldmath$r$},\Omega_{\nu}) 
 \mbox{\boldmath$n$}_{\bar{\nu}} ] \nonumber \\
    && \times \left[ 1-\sin{\theta_{\nu}} \sin{\theta_{\bar{\nu}}}
        \cos{(\varphi_{\nu}-\varphi_{\bar{\nu}})}
        -\cos{\theta_{\nu}} \cos{\theta_{\bar{\nu}}}
    \right]^2
    \sin \theta_{\nu}\sin \theta_{\bar{\nu}}. 
\label{eq:heat_annmom}
\end{eqnarray}
 Finally the remaining procedure is to change the coordinates of the angle integrations  
 in Equations (\ref{eq:heat_ann}) and (\ref{eq:heat_annmom}) from the 
${\bold O}'$ to ${\bold O}$ coordinate system. Albeit sacrificing the formulae to be 
 messy (e.g., appendix A), this procedure, by which 
  the neutrino source and the target can be connected one by one, merits for reducing 
the numerical costs significantly.
 The coordinate transformation with respect to the solid angle is written as, 
\begin{eqnarray}
d\Omega_{\nu} 
		&=& J_{r \mu}(\mu_{\nu}, \phi_{\nu}, r_{s,\nu}, \mu_{s,\nu}) 
			|_{\phi_{s,\nu} = \phi_{{\rm sph},\nu}} dr_{s,\nu} d\mu_{s,\nu} \nonumber \\
		&+& J_{\mu \phi}(\mu_{\nu}, \phi_{\nu}, \mu_{s,\nu}, \phi_{s,\nu}) 
			|_{r_{s,\nu}=r_{{\rm sph},\nu}} d\mu_{s,\nu} d\phi_{s,\nu} \nonumber \\
		&+& J_{\phi r}(\mu_{\nu}, \phi_{\nu}, \phi_{s,\nu}, r_{s,\nu}) 
			|_{\mu_{s,\nu} = \mu_{{\rm sph},\nu}} d\phi_{s,\nu} dr_{s,\nu},
\end{eqnarray}
where ${\mbox{\boldmath$r$}}_{s,\nu} (r_{s,\nu}, \theta_{s,\nu}, \phi_{s,\nu})$ 
is the position of $S_{\nu}$ measured in the ${\bold O}$ system. Using the expressions
 of Jacobian ($J_{r \mu}, J_{\mu \phi}, J_{\phi r}$, see 
appendix A for details), we can calculate the annihilation rates based only on  
 the quantities defined on the hydrodynamical grid-points.

\subsection{Implementation of ray-tracing calculation}
\label{sec:raytrace}

We have shown how to calculate the annihilation rates when neutrinos from the neutrino 
sphere can reach directly to the target region.
However in reality, neutrinos can be absorbed in some 
situations before hitting to the target, for example, by encountering with
 the optically thick region. To correctly count the real contribution, we set
the following three criteria by utilizing the ray-tracing calculation.

The first task is to define an outward neutrino emission from the neutrino sphere.
 This can be satisfied by ${\mbox{\boldmath$b$}}_{\nu} \cdot {\mbox{\boldmath$d$}}_{\nu} >0$, where the former is the normalized neutrino momentum from the neutrino sphere 
${\mbox{\boldmath$d$}}_{\nu} (\equiv {\mbox{\boldmath$p$}}_\nu 
/ |{\mbox{\boldmath$p$}}_\nu| 
 = ({\mbox{\boldmath$r$}} - {\mbox{\boldmath$r$}}_{s,\nu}) 
/ | {\mbox{\boldmath$r$}} - {\mbox{\boldmath$r$}}_{s,\nu}|)$, and the latter is 
the normal direction to the differential surface of the neutrino sphere. 
  As mentioned, we define the surface of the neutrino sphere to be dependent only 
on the lateral direction (e.g., Equation (\ref{mfp})).
 Thus the analytic form of ${\mbox{\boldmath$b$}}_{\nu}$ can 
 be simply given using the valuables such as $\theta_{s,\nu}$ and 
$\phi_{s,\nu}$ that denote the surface position of the neutrino sphere.
Furthermore the criterion of 
${\mbox{\boldmath$b$}}_{\nu} \cdot {\mbox{\boldmath$d$}}_{\nu} >0$ 
would provide more realistic regulation of the neutrino flux than 
 in \citet{ruff97,ruff98} whose criterion is determined only by the density gradient 
 at the neutrino sphere.

The second criterion is 
the positivity of ($2/3 - \bar{\tau}$) along each trajectory from the surface of 
 the neutrino sphere to the target region.
 Here $\bar{\tau}$ denotes the (locally-defined) 
opacity for each neutrino species (e.g., the integrand of the 
left-hand-side of Equation (\ref{mfp})).  
This is necessary to omit the neutrino flux 
which comes from the optically thick region, such as inside the neutrino sphere. 
For not counting radiation absorbing to the black hole (BH), the third criterion is the 
positivity of $r-r_g$, 
 where $r_g = 2GM_{\rm BH}/c^2$ with $G, M_{\rm BH}$ being
 the gravitational constant and the mass of the BH, respectively. 
 For simplicity, we will set $r_g$ to be the Schwartzshild radius 
 when we apply the ray-tracing calculation to the collapsar simulations in the 
later section. However in reality, the BH is rotating so that 
the accurate estimation of the geodesics is necessary, which can be 
 an another possible extension of this study.

\section{Result}
\subsection{Numerical Tests}
\label{sec:result}

Before applying the newly developed code to collapsars, we shall check the accuracy of our code. For the purpose, 
we derive some analytical solutions for test problems and present 
comparison with the numerical results in this section.

Amongst a list of tests, the first requirement we set is to check whether 
the newly derived Jacobians (Equation (23)) 
and their numerical implementations are correct. 
 This will be checked by reproducing radiation fields shedding 
  from a light-bulb in spherical and non-spherical geometry 
(section \ref{subsec:sph_nusph} and  \ref{subsec:sph_bbl}).
The second one is to check the accuracy of our ray-tracing calculation. Note that this can be checked in each test,
 because the ray-tracing is done in whichever tests.
In addition, we present an additional test, which is specially designed to test 
 the validity for the application to the collapsar simulations in 
section \ref{subsec:sph_nobbl}.

In the following tests, 
our numerical domain has the spherical coordinates with equally spaced 100 radial meshes 
 covering from 0 to 500 km in radius. To check the numerical convergence, we
 vary the numerical resolution for $\theta$ and $\phi$ direction, respectively.
For simplicity, the axial and equatorial symmetry is assumed for the hydrodynamic 
 quantities (such as for density, electron fraction and so on) as a background.
 The density and temperature on the neutrino sphere is set to be 
$\rho = 10^{12} \gpcmc$ and $T = 5~{\rm MeV}$, respectively. 

To measure the deviation from the corresponding analytical solutions, 
we estimate the error function of $E$, which is defined by the $L_1$ norm as,
 \begin{eqnarray}
     E &\equiv& \frac{L_{1,{\rm err}}}{L_{1,{\rm ana}}}, \label{eq:err}\\
 	L_{1,{\rm err}} &\equiv& \Sigma | {x_{\rm num} - x_{\rm ana}} | , \\
     L_{1,{\rm ana}} &\equiv& \Sigma {x_{\rm ana}}, 
 \end{eqnarray}
 where
 $x_{\rm num}$ represents the numerical values such as the energy deposition rate 
($q^+_{\nu {\bar{\nu}}}$) and the absolute value of the momentum transfer rate
 ($|\mbox{\boldmath$m$}^+_{\nu {\bar{\nu}}}|$), while
 $x_{\rm ana}$ represents the corresponding analytical solutions.
 $\Sigma$ indicates the summation, which is taken over the whole 
computational domain. 
 
 As for the momentum transfer rate,
 the error function of $E$ cannot always be a useful measure because the 
analytical solution becomes zero in several situations 
(see sections \ref{subsec:sph_bbl} and \ref{subsec:sph_nobbl}). 
 In such a case, we define another error function of $E2$ as,
\begin{eqnarray}
     E2 &\equiv& \frac{M_{\rm num}}{M_{\rm max}}, \\
 	M_{\rm num} &\equiv& \Sigma | \mbox{\boldmath$m$}^+_{\nu {\bar{\nu}}} | , \\
 	M_{\rm max} &\equiv& \Sigma m^+_{\rm max}, 
 	\label{eq:err_mom}
\end{eqnarray}
where $m^+_{\rm max}$ is the analytical solution, which is defined from the energy 
deposition rate as $m^+_{\rm max} \equiv q^+_{\rm ana}/c$, 
thus meaning the maximum momentum transfer. With this $E2$, we will 
evaluate how precisely the cancellation can be maintained.

\begin{figure}[tbd]
\epsscale{.8}\plotone{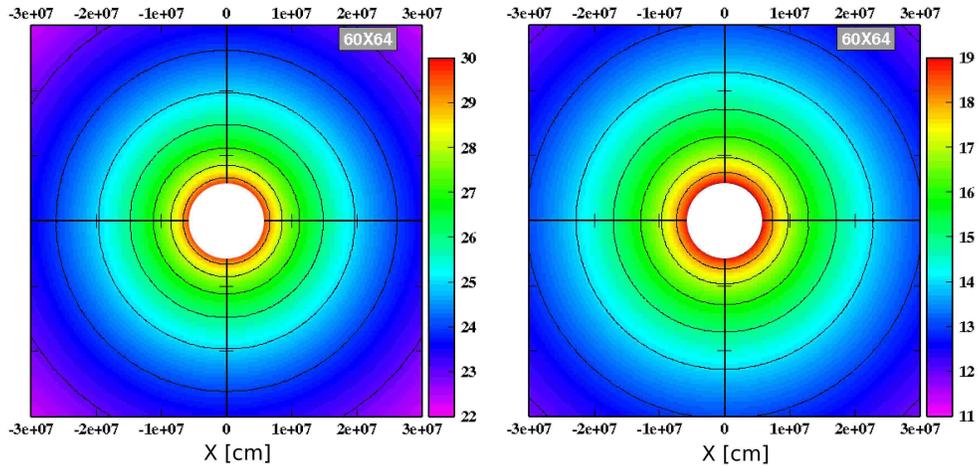}
\caption{
Energy deposition (left panel) and momentum transfer rates (right panel)   
from a spherical neutrino sphere (central white circle).
 In each panel, left-half and right -half represents the numerical and analytical 
solutions, respectively. 
The numerical resolution is set to be 
 $(n_\theta, n_\phi) = (60, 64)$, 
 which are indicated in the top right edge in 
 each panel as ``60X64''. 
With Figure 4, our code is shown to 
reproduce the sphericity of the radiation fields well.}
\label{fig:nusph}
\end{figure} 
\clearpage
\subsubsection{Radiation from Spherical Neutrino Sphere}
\label{subsec:sph_nusph}

As a simplest test, we first present how accurately we can reproduce the 
radiation field from a spherical source, which emits radiation 
isotropically from its surface. In this test, we set a spherical inner boundary of 
50 km in radius, 
 inside which material is taken to be completely 
optically thick for neutrinos, while outside is 
 taken to be completely optically thin. Albeit very crude, such a situation is akin 
 to the neutrino radiation from the neutrino sphere 
in the postbounce phase of core-collapse supernovae. 
 In addition to this static source, we consider another case in which the 
 spherical neutrino sphere rotates relativistically. By this test, we hope to see and
 test the special relativistic effects, 
which is one of the main focus in this paper. 

\begin{figure}[tbd]
\epsscale{1}\plotone{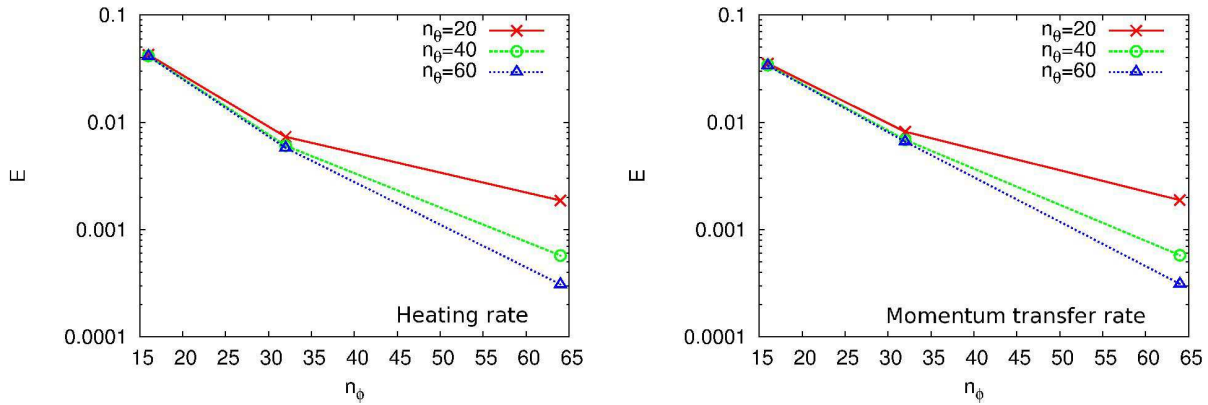}
\caption{Deviation from the analytical solutions for the energy (left) and momentum 
 transfer rates (right) as a function of $n_{\phi}$ and $n_{\theta}$ 
in the case of the spherical neutrino sphere.
 See Equation (\ref{eq:err}) for the definition of error $E$, which is calculated by the $L_1$ norm.}
\label{fig:err_nusph}
\end{figure}

\subsubsubsection{Non Relativistic Case}
\label{subsubsec:sph_nusph_nosr}

In the case of the static neutrino sphere, the analytic formulae of the 
 heating and momentum transfer rate have been explicitly given 
 in \citet{good87,coop87,salm99,asano2}, as 
\begin{equation}
    \frac{dq_{\nu {\bar{\nu}}}^+(\mbox{\boldmath$r$})}{dt dV}
	= \frac{  4 c K G_{\rm F}^2 }{3}
\frac{(kT)^9}{(hc)^6}{\cal F}_4(0){\cal F}_3(0) F(r), 
\label{eq:nusph_nosr}
\end{equation}
and 
\begin{equation}
    \frac{d{\bold m}_{\nu {\bar{\nu}}}^+(\mbox{\boldmath$r$})}{dt dV}
	= \frac{  4 K G_{\rm F}^2 }{3}
\frac{(kT)^9}{(hc)^6}{\cal F}_4(0){\cal F}_3(0) G(r) {\mbox{\boldmath$n$}}_r, 
\label{eq:nusph_nosr_mom}
\end{equation}
where geometrical factor $F(r)$ and $G(r)$ is given by
\begin{equation}
F(r) = \frac{2\pi^2}{3} (1-X)^4 (5 + 4X + X^2), 
\label{eq:geo_heat}
\end{equation}
and 
\begin{equation}
G(r) = \frac{\pi^2}{6} (1-X)^4 (1+X)(8 + 9X + 3X^2), 
\label{eq:geo_mom}
\end{equation}
 with 
\begin{equation}
X = \sqrt{1-\left( \frac{R_\nu}{r}\right)}, 
\end{equation}
where $R_{\nu}$ is the radius of the neutrino sphere.

Figure \ref{fig:nusph} shows 
the energy (left panel) and momentum (right panel) rates 
numerically calculated with our code 
(left-half) and 
the corresponding analytical solutions from Equation (\ref{eq:nusph_nosr}) 
(right-half). 
Note that the label such as "20X16" at the top right edge indicates the
  resolution of our ray-tracing calculation, here 
$n_\theta = 60, n_\phi = 64$,
 the equally spaced mesh numbers in the lateral ($\theta$) and azimuthal direction
 ($\phi$), respectively. Comparing the 
two panel, we find no visible differences in each panel.
 In Figure \ref{fig:err_nusph}, 
the deviation from the analytical solution for the different numerical resolutions 
(e.g., Equation (\ref{eq:err})) is shown.  As shown, the relative error decreases with 
$n_\theta, n_\phi$ as low as 0.1 \%. Compared to other studies treating the 
radiation problems (e.g., \citet{ston92,turn01}), the obtained accuracy here seems 
high enough. These results assure the validity both of the newly derived expression of 
$J_{\mu \phi}$ (from Equation (\ref{eq:jac_start}) to (\ref{eq:jac_end})) 
and its implementation. 

\begin{figure}[tbd]
\epsscale{.8}\plotone{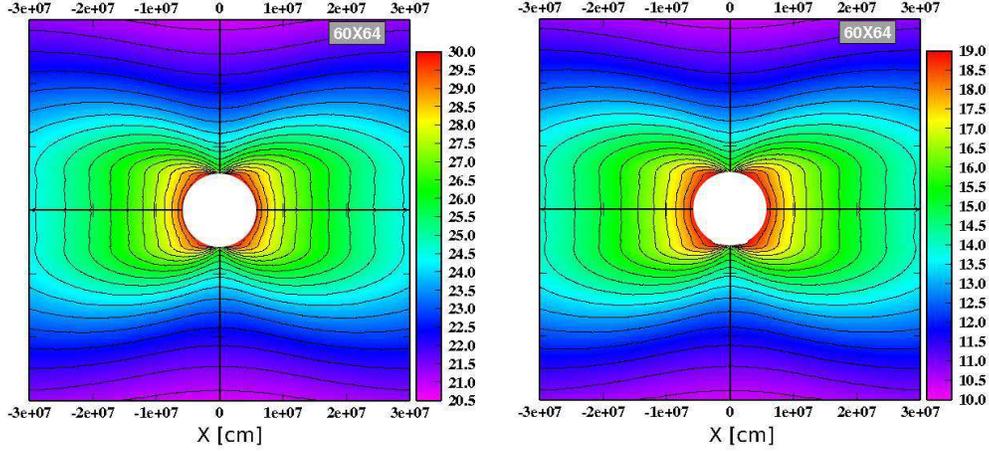}
\caption{Energy deposition rate (left) and radial momentum transfer rate (right) 
from the relativistically rotating neutrino sphere (central white circle). 
Comparing with Figure \ref{fig:nusph}, the energy and momentum 
 rates are shown to be significantly reduced along the rotational axis 
by the beaming effect. (t is noted that the numerical resolution taken 
 here is $(n_\theta, n_\phi) = (60,64)$. 
}
\label{fig:nusph_sr}
\end{figure} 

\subsubsubsection{Relativistic Case}
\label{subsubsec:sph_nusph_sr}

For the neutrino sources with relativistic motion, it is not generally possible
 to write down the special relativistic factor of $\xi_{\nu}$ (Equation (10)) 
in an analytic form. This is because it depends locally 
on the angle between the velocity field and the flight direction of 
 neutrinos. To see special relativistic effects, we examine a special case, 
in which the analytic solution can be found.
 The annihilation rates along the polar axis, emitted from the relativistically 
{\it rotating} neutrino source, can be analytically given as, 
\begin{eqnarray}
    \frac{dq_{\nu {\bar{\nu}}}^+(\mbox{\boldmath$r$})}{dt dV}
	&=& \frac{  4 c K G_{\rm F}^2 }{3}
\frac{(kT)^9}{\gamma^9 (hc)^6}{\cal F}_4(0){\cal F}_3(0) F(r), 
	\label{eq:nusph_sr} \\
    \frac{d{\bold m}_{\nu {\bar{\nu}}}^+(\mbox{\boldmath$r$})}{dt dV}
	&=& \frac{  4 K G_{\rm F}^2 }{3}
\frac{(kT)^9}{\gamma^9 (hc)^6}{\cal F}_4(0){\cal F}_3(0) G(r) {\mbox{\boldmath$n$}}_r, 
	\label{eq:nusph_sr_mom}
\end{eqnarray}
where the geometrical factors $F(r)$  and $G(r)$ are 
just same as Equations (\ref{eq:geo_heat}) and (\ref{eq:geo_mom}), respectively. 
 The sharp suppression with a factor of $1/\gamma^9$ is the outcome of the 
 relativistic beaming, simply because the rotational velocity is perpendicular to the 
polar direction. We will see it soon in the following.

\begin{figure}[tbd]
\epsscale{.8}\plotone{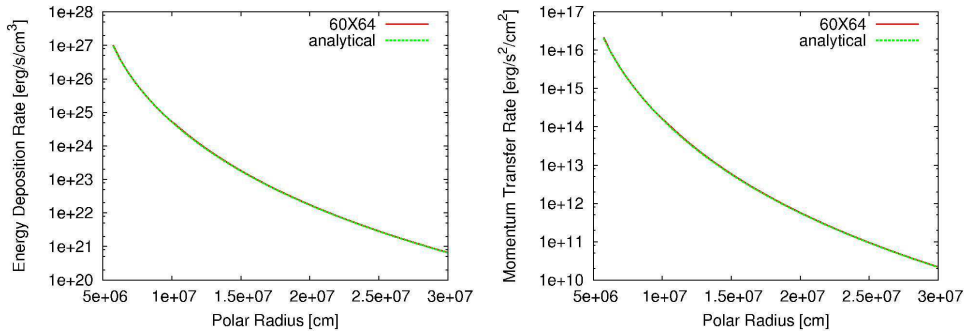}
\caption{Comparison between the numerical (solid line) and analytic (dashed line)
 solution of the energy (left) and momentum transfer rates (right) 
for the case of the relativistic rotating neutrino sphere. 
It is noted that the numerical resolution taken 
 here is $(n_\theta, n_\phi) = (60,64)$. }\label{fig:sph_sr_pol}

\end{figure}

\begin{figure}[tbd]
\epsscale{.8}\plotone{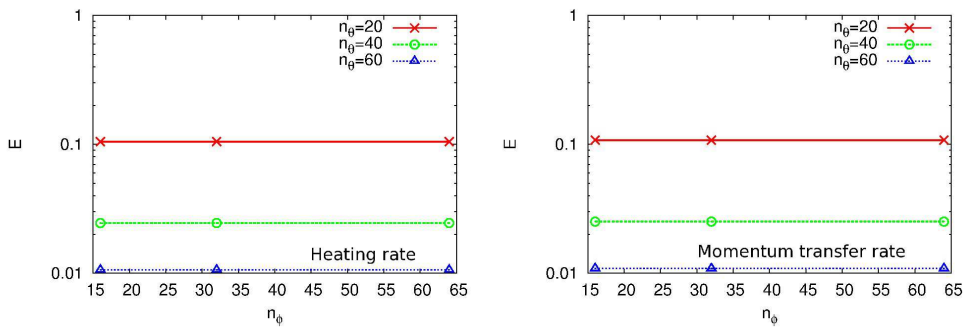}
\caption{Same as Figure \ref{fig:err_nusph} 
but for the case of the relativistic rotating neutrino sphere. }
\label{fig:err_nusph_sr}
\end{figure}

Figure \ref{fig:nusph_sr} shows 
the energy deposition (left) and momentum transfer rates (right) from the 
 relativistically rotating neutrino sphere.
Here we assume $\gamma = 2$ with vanishing $V_r$ and $V_{\theta}$.
Note in this figure that the numerical results are only shown, 
owing to the inaccessibility of the analytical solution as mentioned above.
 It is shown that the annihilation rates both for energy and momentum
 is greatly reduced near on the rotational axis as expected from Equations (\ref{eq:nusph_sr}) and (\ref{eq:nusph_sr_mom}), 
while they are enhanced around the equatorial plane. 
Figure \ref{fig:sph_sr_pol} shows 
the comparison of the numerical solution (solid line) with the 
analytical solution (dashed line) along the rotational axis for energy
(left) and momentum (right) transfer rates. 
In support of the validity of our special relativistic implementation, 
no visible differences are seen between the analytic and numerical solutions. 
 Figure \ref{fig:err_nusph_sr} shows the deviation of $E$ (e.g., Equation (\ref{eq:err})) 
for the energy (left) and 
momentum (right) from the analytic solution. From this test, it is shown that 
the lateral grid points ($n_{\theta}$) 
should be at least more than 20 to maintain the percent levels of 
agreement with the analytic solution.


\subsubsection{Spherical Cavity}
\label{subsec:sph_bbl}

In this test, we set a spherical cavity of $300$ km in radius,
 outside which is the uniform opaque neutrino source. 
 Although this is not a realistic astrophysical situation, 
 it provides one of the most simplest null test 
 of the momentum transfer because the neutrino flux is isotropic everywhere 
inside the cavity.  For the heating rate, this test merits because the geometrical factor
(Equation (\ref{eq:nusph_nosr})) is analytically given by 
\begin{equation}
F(r) = \frac{64 \pi^2}{3}.
\label{eq:sph_bbl}
\end{equation}

\begin{figure}[tbd]
\epsscale{1.}\plotone{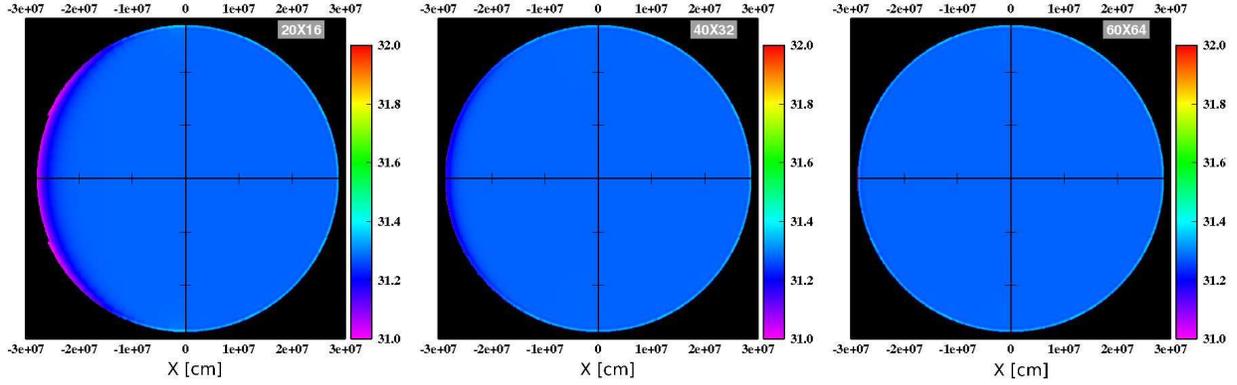}
\caption{Same as Figure \ref{fig:nusph} but for the energy deposition rates 
 in the case of the spherical neutrino cavity.}
\label{fig:nusph_i}
\end{figure}

\begin{figure}[tbd]
\epsscale{1.}\plotone{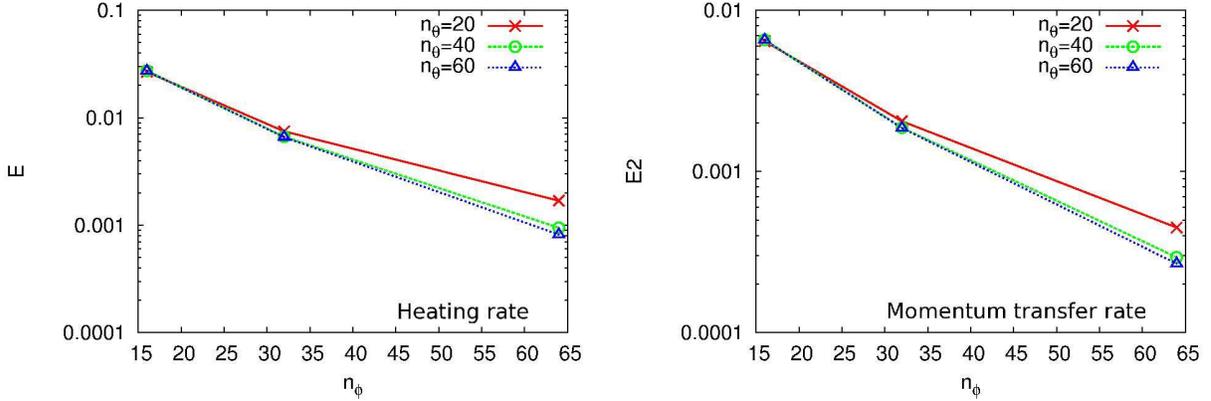}
\caption{Same as Figure \ref{fig:err_nusph} 
but for the test of the spherical neutrino cavity.
The error of $E2$ (right) is calculated with the 
$L_1$ norm, which is defined in Equation (\ref{eq:err_mom}). }
\label{fig:err_nusph_i}
\end{figure}

Comparing the numerical (left-half) and analytical solution (right-half) in each 
 panel of Figure \ref{fig:nusph_i}, the discrepancy is shown to become 
smaller for the better numerical resolution (from left to right). 
 From Figure \ref{fig:err_nusph_i} showing the $E1$ and $E2$ errors
 for the energy (left) and momentum transfer rate (right), it can be seen
 that the errors mainly depend on $n_\phi$. By setting  $n_\phi = 32$,
  we can get the agreement with an accuracy of 1 \% for $E$ and 
 0.1 \% for $E2$, which would satisfy the requirement of 
the vanishing momentum transfer sufficiently.


\subsubsection{Non-Spherical Cavity}
\label{subsec:sph_nobbl}

We now move on to the non-spherical cavity test, by which we can check if the
 radiation contribution from the face element of $drd\phi$ is accurately estimated.
For simplicity, we just modify the shape of the spherical cavity in the previous section.
We set the position of the neutrino sphere to change with the polar angle $\theta$. 
For $\theta < \pi/4$ and $\theta > 3 \pi/4$, the neutrino sphere is situated at 300 km, 
while for $\pi/4 \le \theta \le 3 \pi/4$, at 100 km. 
Region inside the neutrino sphere is taken to be completely thin, 
while completely thick outside. Even with the geometrical 
difference, it should be noted that the items to be checked are the same 
 for the case of the spherical cavity, because the net flux is isotropic 
 and does not change.

\begin{figure}[tb]
\epsscale{1}\plotone{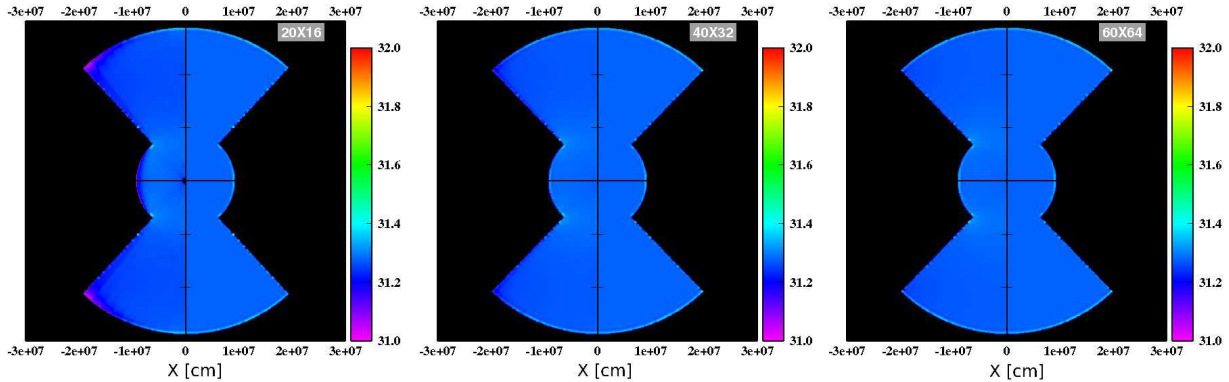}
\caption{Same as Figure \ref{fig:nusph} but for the energy deposition rates 
 in the case of the non-spherical neutrino cavity.}
\label{fig:nusph_i_a}
\end{figure}

\begin{figure}[tb]
\epsscale{1}\plotone{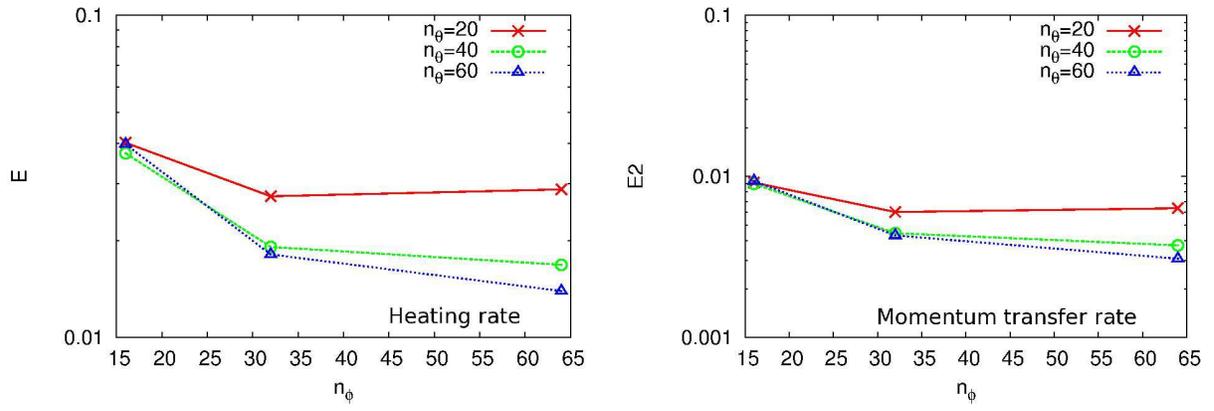}
\caption{Same as Figure \ref{fig:err_nusph_i} but 
for the case of the non-spherical neutrino cavity. }
\label{fig:err_nusph_i_a}
\end{figure}

From Figure \ref{fig:nusph_i_a}, it can be seen that some inhomogenities 
 seen closely at the edge of the boundary (left-half: numerical) 
become smaller for the higher resolutions (to the right panel).
 From Figure \ref{fig:err_nusph_i_a},  the decreasing rate of the 
 relative errors for $E$ and $E_2$ is shown to be smaller than the ones from the 
previous tests due to the relatively complicated geometry. 
By taking the resolution of ($n_\theta$, $n_\phi) \gtrsim (32,32)$, our code can 
 reproduce the analytic solutions with percent levels of accuracy here. 

 As evident from the numerical convergence with increasing the numerical 
resolution (e.g., see Figures 4, 7, 9, and 11), the previous tests have 
supported the validity of the newly derived expression of the Jacobians
 (from eq.(\ref{eq:jac_start}) to (\ref{eq:jac_end})), its 
implementation, and the ray-tracing calculation.  From those test calculations, 
 it is suggested that the grid numbers 
of $(n_\theta, n_\phi) \gtrsim (30, 30)$ are required for obtaining the percent levels
 of agreement with the analytical solution. Considering the computational cost, 
we choose to employ the grid numbers of 
$(n_r, n_\theta, n_\phi) = (300, 40, 32)$ in the actual implementation for 
 collapsar simulations in the following section. 


\clearpage

\subsection{Application to Collapsar}
\label{sec:application}

Having demonstrated the accuracy of our code in the previous sections, 
 we now proceed to show an application to collapsars.
In this section, we estimate the annihilation rates in 
a post-processing manner based on our simulation results of the long-term evolution
 of collapsar and then discuss the possible impacts on the collapsar dynamics. 

\subsubsection{Recipes for Collapsar Simulation}
 For obtaining the hydrodynamic profiles for 
 the ray-tracing calculation, we perform the special relativistic simulations of 
collapsar  \citep{hari09}.
 Without repeating the numerical procedures again,
 we will shortly summarize major modifications added in this work,
 namely about the initial model and 
 the position of the inner-boundary.

 Since our focus is not on the MHD-driven collapsars,
 we construct initial models with no magnetic fields.
 Taking the precollapse density/electron-fraction/entropy profiles of progenitor of 35OC in \citet{woos06}, we embed 
the angular velocity $\Omega(r,\theta)$ analytically as 
\begin{equation}
 \Omega(r,\theta)=\frac{\Omega_0 X_0^4 + \alpha \Omega_\mathrm{lso}(M(X))X^4}{X_0^4+X^4}, 
 \label{eq:iniang}
\end{equation}
where $X=r\sin\theta$, 
$\Omega_\mathrm{lso}(M(X)) = j_{\rm lso}(M(X)) / X^2 $. 
$M(X)$ is the mass coordinate at $X$, and 
$j_{\rm lso}$ and $\Omega_\mathrm{lso}$ is the specific angular momentum and angular 
 velocity in the last stable orbit. 
Model parameters are $\alpha, \Omega_0,$ and $X_0$. 
This distribution provides 
co-rotation with $\Omega = \Omega_0$ at $X < X_0$ connected to 
constant specific angular momentum 
$j = \alpha j_{\rm lso}$ at $X > X_0$. 
$X_0$ would correspond to the size of co-rotating region in pre-collapse phase, 
such as Fe core. To be closer to the original rotational profile in \citet{woos06},
we set $\alpha = 0.8$, 
$\Omega_0 = 1.2 \radps$, and 
$X_0=R_{\rm Fe}$, 
where $R_{\rm Fe}$ is the size of Fe core ($\approx 3000 \km$ for 35OC).

Another major modification is the position of the absorbing inner boundary of 
 the computational domain. For reducing the 
computational cost, the radius of the boundary, which mimics the surface of black hole, 
was set to be $50$ km \citep{hari09}. 
This manipulation may lead to the underestimation of the neutrino 
luminosity by excising the radiation from more inner edge of 
the accretion disk.  This is obviously disadvantageous for 
 producing the neutrino-driven outflow. In this paper, we take more compact
inner-boundary as $max (10 \km ,2r_g)$, extending to the outer boundary of 30000 km. 

\begin{figure}[tbd]
\epsscale{1.0}\plotone{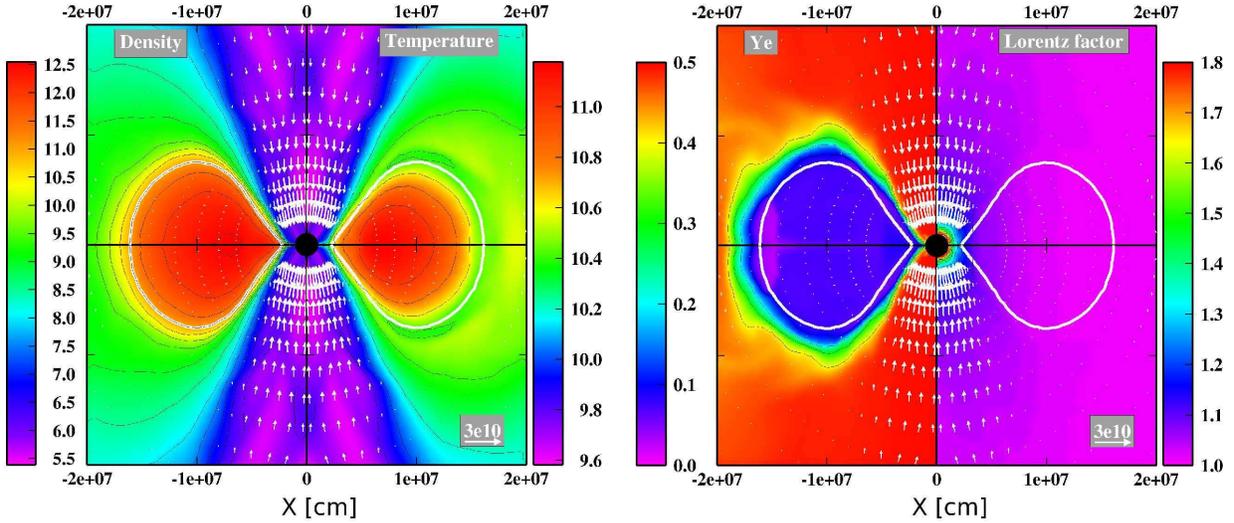}
\caption{Typical hydrodynamic configuration when the accretion disk is in 
a stationary state (here, 9.1 s after the onset of gravitational collapse). 
In the left panel, the logarithmic density (in$\gpcmc$, left-half) and
 temperature (in $K$, right-half) are shown. 
The right panel shows the electron fraction (left-half)
 and the Lorentz factor (right-half). 
The velocity, indicating by white arrows, is normalized 
by the scale in the bottom right edge of the box (here, $3\times10^{10} \cmps$). 
 The white solid line denotes the area where the density is equal to $10^{11} \gpcmc$,
 representing the surface of the accretion disk. 
The central black circle (11.5 km in radius ($\approx 2 r_g$)) represents
the inner boundary of our computations. 
}
\label{fig:hydro}
\end{figure} 

\subsubsection{When can the pair neutrino annihilation work ?}
For energizing the jets via pair neutrino annihilation, the higher neutrino 
luminosity from the accretion disk and the lower density along the polar region,
 are favorable. Our collapsar model starts with a rapid mass accretion into
 the central objects. Within $\sim$ 2.0 s after the onset of gravitational collapse, 
 this model accretes more than 2$\sim 3 \Ms$ in the center. This indicates
 the black hole formation in the center because the maximum mass of the neutron star,
 which the Shen equation of state employed here can sustain, is in the same 
 mass range \citep{kiuchi08}. 
Simultaneous with the rapid infall, the neutrino luminosities also show 
a drastic increase, and then shift to a gradual increase reflecting 
the mass accretion to the newly formed accretion disk. 
At $\sim $ 8 s from the onset of gravitational collapse, 
the total neutrino luminosities gets as high as $10^{53} \ergps$, which consists of 
$2.2 \times 10^{52} \ergps$ for $\nu_{\rm e}$, 
$9.4 \times 10^{52} \ergps$ for ${\bar \nu}_{\rm e}$, 
$5.1 \times 10^{51} \ergps$ for $\nu_{\mu}$ and $\nu_{\tau}$.  
 Afterwards, the total luminosity keeps almost constant with time, reflecting that the 
 accretion disk comes to a (quasi)stationary state. 

 Figure \ref{fig:hydro} shows the distributions of density, temperature, 
electron fraction, and Lorentz factor of our model at 9.1 s after the onset of 
gravitational collapse.
From the left panel, the structure of the accretion disk with a polar funnel along 
 the rotation axis is shown. The disk temperature reaches typically $\sim 10$ MeV, 
 which provides the high neutrino luminosity of $10^{53} \ergps$.   
 From the right panel, the Lorentz factor at the inner edge of the disk becomes 
$\gamma \approx 1.3$, indicating the importance of the special relativistic effects.
 Taking these hydrodynamical data, we estimate
the neutrino pair annihilation by the ray-tracing calculation in a 
post-processing manner.

\subsubsection{Energy deposition and momentum transfer rates in collapsar}
\label{subsubsec:ene_mom}

\begin{figure}[tb]
\epsscale{1.0}\plotone{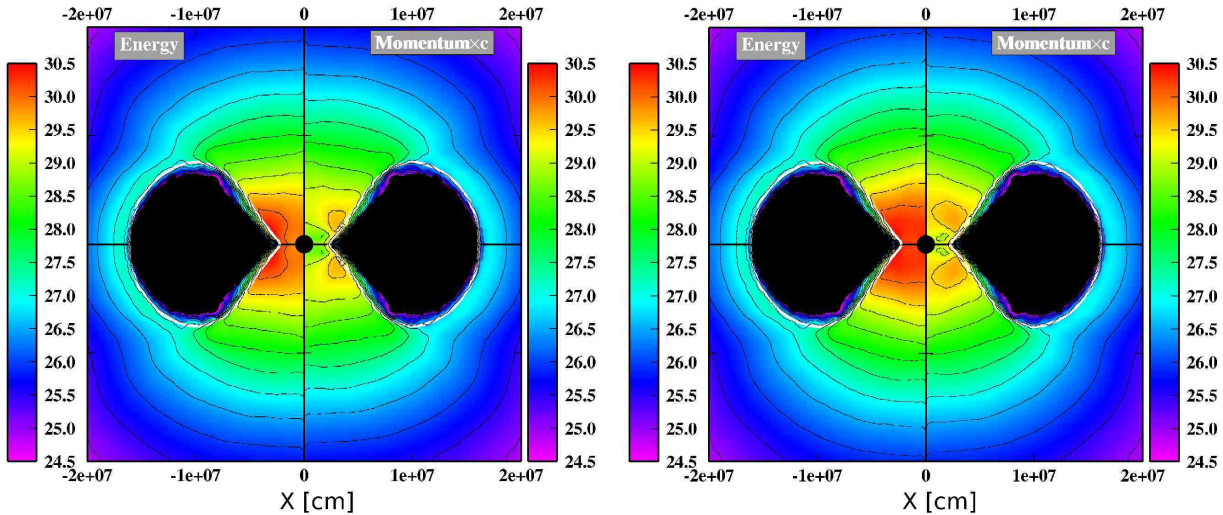}
\caption{Same as Figure 12 but for the energy and momentum transfer rates with 
(left) and without special relativistic corrections (right). 
In each panel, logarithmic contour of 
the energy deposition rate (left-half) and 
the momentum transfer rate multiplied by the speed of light (right-half) 
 both in unit of (${\rm erg}~{\rm s}^{-1}~{\rm cm}^{-3}$) are drawn. }
\label{fig:qmom}
\end{figure}

The left- and right- hand sides of the left panel of Figure \ref{fig:qmom} shows 
the energy deposition and the momentum transfer rates (its absolute value). 
 The energy and momentum transport from 
 the accretion disk (regions colored by black) to the polar funnel regions (regions 
 colored by red) can be apparently seen. 
Interestingly, the transfer rates become 
 highest near the inner edge of the accretion disk.
 Inside $\sim$ 80 km in radius from the center, 
the typical energy deposition rates 
along the polar axis are $10^{30}~{\rm erg \,\, s^{-1} \,\, cm^{-3}}$, where the 
momentum transfer rates are the order of $10^{19}~{\rm erg \,\, cm^{-4}}$.
 Thus the momentum transfer rates multiplied by the speed of light $c$ becomes 
comparable to the energy deposition rate there. 

For our model, the density along the polar funnel regions is as low as 
$10^{5-6} \gpcmc$ (see Figure \ref{fig:hydro}(left)). 
Assuming that the momentum transfer there is maintained for 10 ms with 
conserving its momentum, we can estimate that the material along the funnel 
can be accelerated up to the Lorentz factor of $10^{1-2}$. 
 This suggests that the 
momentum transfer plays as important role as the energy deposition for the 
 efficient acceleration of the neutrino-driven outflows.
 The right panel of Figure \ref{fig:qmom} is the annihilation rates without the special relativistic correction ($\xi_{\nu} = 1$ in Equation (\ref{eq:heatdef_srstart})).
 Comparing the two panels in Figure \ref{fig:qmom},
we find that the special relativistic effect decreases the transfer rates 
by several factors inside the polar funnel. 
As already discussed in section \ref{subsec:sph_nusph}, this 
 is the outcome of the relativistic beaming by the rapidly rotating accretion disk.
  On the other hand, the reduction in the 
total deposition rates, given by the volume integral of the local rates 
 in Figure \ref{fig:qmom}, can stay relatively smaller, namely 
of  
$6.22$ and $7.62/(10^{50} \ergps)$
with and without the correction, respectively. 
These values are comparable to what expected by \citet{naga03}, 
since the mass accretion rate in this epoch is about $0.1 \Ms\psec$. 
 The resulting conversion efficiency of the heating, which is defined by the 
 ratio of the total deposition rates to the total neutrino luminosity, is 
0.514\% and 0.630 \%
with and without the correction. 
Those numbers are comparable to the 
ones in the previous studies (e.g., \citet{ruff97,ruff98,ruff99,nara01,seti04,seti06}), 
and this negative effect of special relativity on the energy deposition rate 
is also mentioned by \citet{birkl}.

\begin{figure}[tb]
\epsscale{1.0}\plotone{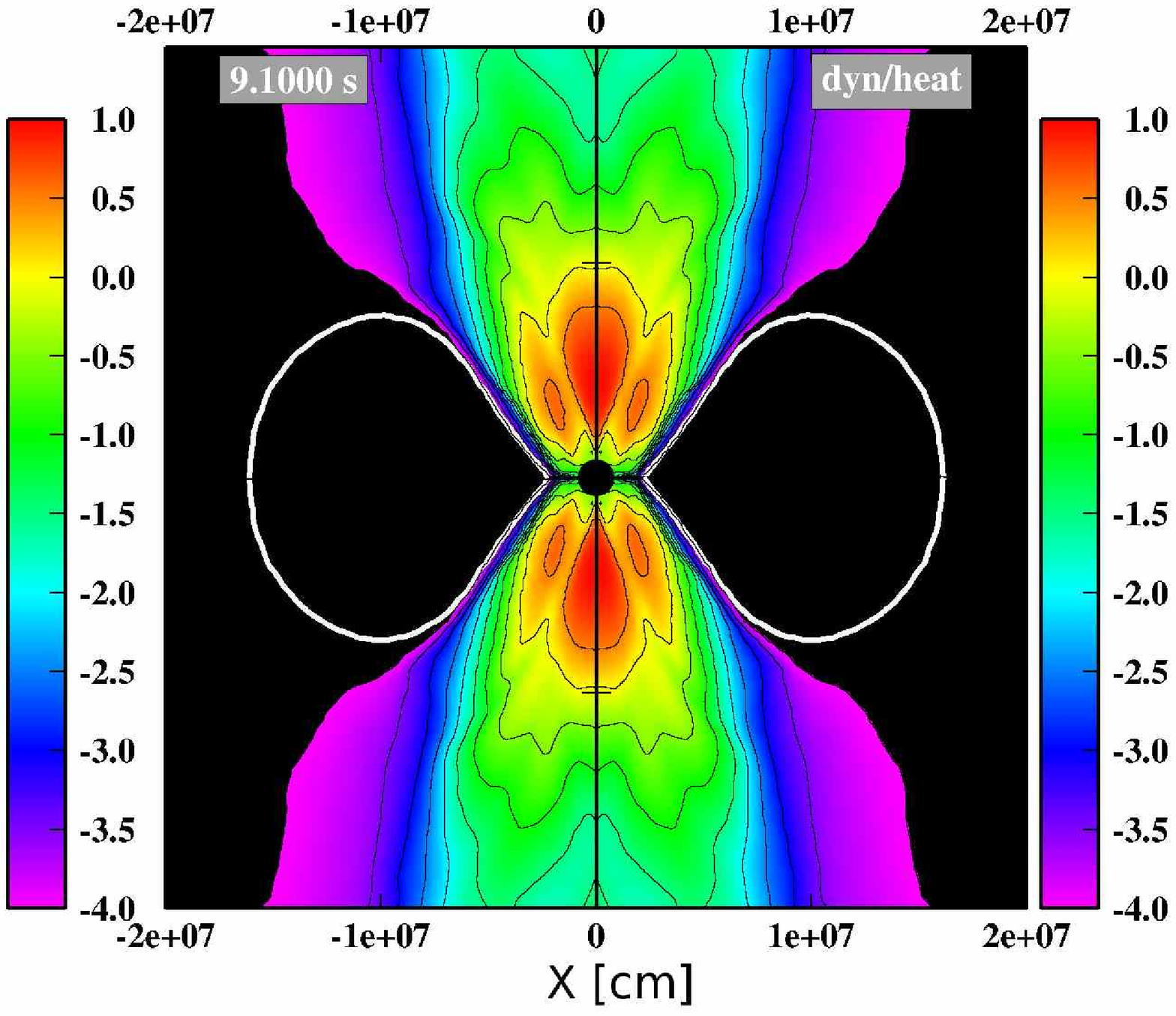}
\caption{Same as Figure \ref{fig:qmom} 
but for the logarithmic contour of 
the dynamical timescale $\tau_{\rm{dyn}}$ divided by 
the heating timescale $\tau_{\rm{heat}}$. This is at the epoch of 
 9.1 s after the onset of gravitational collapse, 
when the neutrino luminosity is as high as $10^{53} \ergps$. }
\label{fig:ht}
\end{figure}

\subsubsection{Comparison of timescales}
\label{subsubsec:timescale}

Based on the annihilation rates in the last section, we compare various timescales in 
 this section, such as the neutrino-heating timescale and the dynamical timescale. 
Then we anticipate if the neutrino-heating outflows could be produced or not in 
the polar funnel regions.

To trigger the neutrino-heating explosion, the neutrino-heating timescale 
should be smaller than 
the advection timescale, which is characterised by the free-fall timescale in the 
 polar funnel regions. 
This condition is akin to the condition of the successful neutrino-driven explosion
 in the case of core-collapse supernovae 
(e.g., \citet{bethe} and see collective references in 
\citet{janka07}).
 The heating timescale is the timescale 
 for a fluid to absorb the comparable energy to its gravitational binding energy to be 
 gravitationally unbound, which is defined as $\tau_{\rm{heat}} \equiv \bar{\rho} \Phi / q^+$, where $\bar{\rho}$ is the 
average density at a certain radius and we take $\bar{\rho}(r) \equiv 3 M(r) /4 \pi r^3$,
$\Phi$ is the local gravitational potential. 
The dynamical timescale is defined as $\tau_{\rm{dyn}} \equiv \sqrt{3\pi/16G\bar{\rho}}$.
  Figure \ref{fig:ht} depicts the ratio of the dynamical $\tau_{\rm{dyn}}$ to 
the heating timescales $\tau_{\rm{heat}}$, showing that the ratio becomes 
greater than unity inside 80 km in the vicinity of the rotational axis. This
 indicates the possible formation of the
 neutrino-driven outflows, if coupled to the collapsar's hydrodynamics.

\begin{figure}[tb]
\epsscale{1.0}\plotone{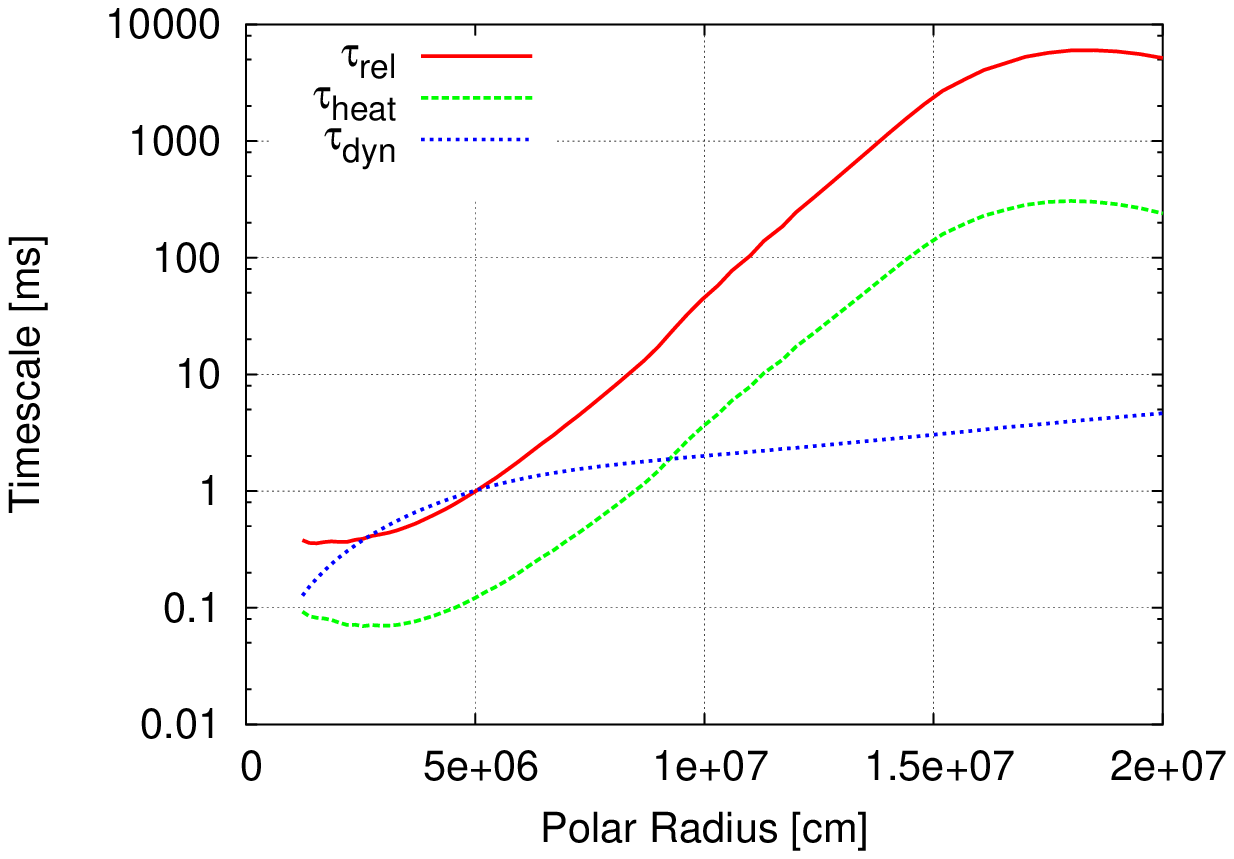}
\caption{Plots of various timescales for material along the rotational axis 
in Figure 14 (see text for the definition of the timescales).}
\label{fig:timescale}
\end{figure}

Figure \ref{fig:timescale} shows various timescales for the material
 along the rotational axis in Figure  \ref{fig:ht}. 
$\tau_{\rm rel} \equiv \rho c^2 / q^+$ characterizes the 
timescale for matter to become relativistic by the neutrino heating.
As mentioned, the first criterion of $\tau_{\rm dyn} > \tau_{\rm heat}$, 
is satisfied for the regions inside $\sim$ 80 km (compare green and blue lines). More interestingly, inside $\sim$ 50 km, $\tau_{\rm rel}$ gets slightly shorter than $\tau_{\rm dyn}$ (compare red 
and blue lines), indicating that matter could attain relativistic motions there.
 To see these outcomes needs the implementation of the ray-tracing calculation 
in the hydrodynamic calculation, which we pose as a next task of this study.
It is noted that with our choice of the grid numbers 
($(n_r, n_\theta, n_\phi) = (300, 40, 32)$), it generally takes several minutes 
(in the CPU time) for each ray-tracing calculation, using 256 processors 
in the Cray XT4 system at the National Astronomical Observatory in Japan.
To follow the dynamics typically later than 10 s after the 
onset of core-collapse for satisfying the $\tau_{\rm dyn} > \tau_{\rm heat}$
 condition, 10000000 hydro-steps are generally needed because the hydrodynamical 
timescale is an order of $10^{-6}$ s in our simulation. 
Therefore it is still computationally expensive to turn on the ray-tracing 
calculation throughout the whole simulation. In the actual implementation,
 we plan to switch on the ray-tracing calculation 
by monitoring intermittently $\tau_{\rm dyn} / \tau_{\rm heat}$  during the simulations 
in the computational domain. After the ray-tracing calculation is turned on, 
seeing $\tau_{\rm dyn} / \tau_{\rm heat} \gtrsim 1$,
 the neutrino-driven outflows will be launched within 0.1 s 
(Harikae et al. in preparation). Thus the computational time using the above 
supercomputer can be reduced to several months, which are expected to be quite feasible.

\begin{figure}[tb]
\epsscale{1.0}\plotone{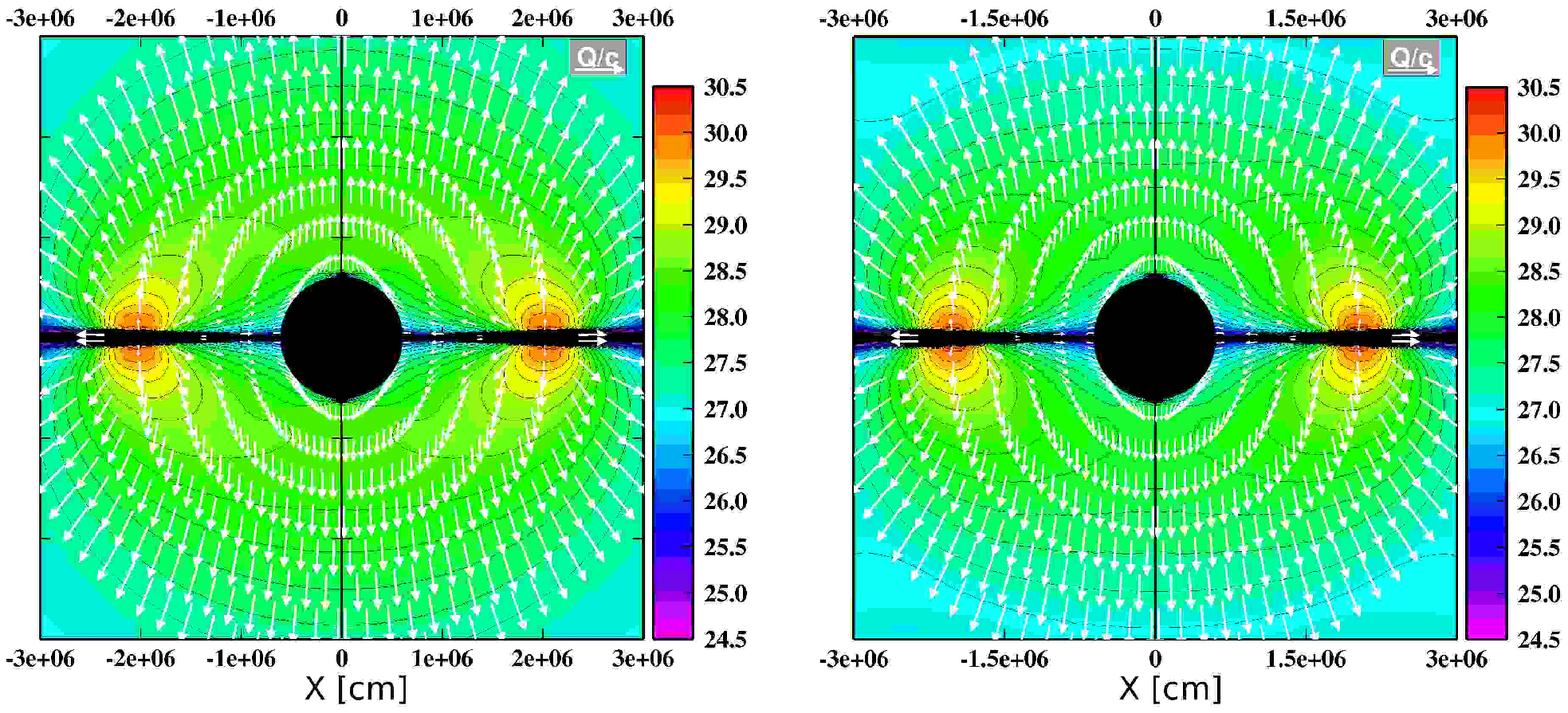}
\caption{Annihilation rates for the thin disk models in \citet{birkl} (for model DN (left)
 with the Newtonian ray-tracing and for model DN4L with the special relativistic one 
(right). The value of the energy deposition rate ($Q^{E}$, e.g., Equation (9)) 
is color-coded, while the spatial 
 vector is visualized by showing the spatial velocity vector $v \equiv Q^{M}/Q^{E}$
($Q^{M}$ is equal to Equation (22)), which is normalized by the speed of light $(c=1)$
 being represented by the arrow (top right in each panel).
A good agreement with \citet{birkl} is obtained by comparing to their Figure 2 (top right).}
\label{fig16}
\end{figure}

\section{Summary and discussion}
\label{sec:discussion}
Bearing in mind the application to the collapsar models of gamma-ray bursts,
we have developed a numerical scheme and code for calculating the energy and momentum 
transfer via neutrino pair annihilation in the framework of special relativity.
To estimate the transfer rates, 
we perform a ray-tracing calculation of the neutrino flux from the accretion disk
 of collapsars. We have tested the numerical accuracy of the developed code 
by several numerical tests, in which the comparisons with the corresponding
 analytical solutions were shown.  Using hydrodynamical data obtained in our 
collapsar simulation, we have run our code 
to estimate the annihilation rates in a post-processing manner.
 We have obtained the following results.

For the progenitor chosen in this study,
the accretion disk settles into a stationary state typically 
$\sim 9$ s after the onset of gravitational collapse. The accretion disk later 
 on can be luminous source of neutrinos as high as $\sim 10^{53} \ergps$. 
At this epoch, the 
  polar funnel along the rotation axis is formed, which is heated intensively 
 from the disk via neutrino pair annihilation. The beaming 
effects of special relativity are found to 
decrease the transfer rates by several factors 
in the polar funnel regions. Inside 
$\sim$ 80 km in the vicinity of the polar funnel, 
we point out that the neutrino-heating timescale 
can become shorter than the hydrodynamical timescale. This indicates the 
 possible formation of the neutrino-heated outflows there.
Our results also suggest that the 
 momentum transfer via neutrino pair annihilation 
plays as important role as the energy deposition for the 
 efficient acceleration of the neutrino-driven outflows.

Finally we shall make comparison with previous work and also mention 
some imperfections of this study. As mentioned, 
 general relativistic effects ignored in this study have been considered as 
an important ingredient of the neutrino pair annihilation in collapsars 
\citep{jaros93,jaros96,salm99,asano1,asano2,birkl}.
It is noted that we have calculated the energy and momentum deposition 
 rates for the thin disk models in \citet{birkl} to assess the validity of our code.
 Our code can reproduce their results well both for the Newtonian and special 
 relativistic case (see Figure \ref{fig16} for the 2D configurations). 
In fact, the total heating rates are 
$Q^{\rm tot} = 2.0\times 10^{48}$ and $1.5\times 10^{48} (\rm{erg}~\rm{s}^{-1})$ 
for the Newtonian (model DN) and the special relativistic case (model DN4L), 
which are in \citet{birkl}, $1.5 \times 10^{48}$ and 
$1.3 \times 10^{48} (\rm{erg}~\rm{s}^{-1})$, respectively 
(see Table 1 in \citet{birkl}). 
When we boldly extrapolate results in \citet{birkl} to ours, 
the local heating rates can be enhanced significantly (by several factors) 
 in the vicinity of the central BH due to the GR effects.
 To update the present scheme to the one in general relativity
 requires
 not only the ray-tracing in the curved space, but also the general relativistic 
formulation of radiative transport, which we are to investigate
 as extension of this study. 
At the same time, we should improve the ray-tracing calculation to 
take into account the charged-current absorption occurring along a given ray, 
which is ignored in this study. Without the neutrino absorption, 
the heating rates estimated in this paper should give a upper bound.
We also plan to include the effects of magnetic
 fields on this study. By changing the strength of magnetic fields systematically,
 we hope to clearly understand how the outflow production in collapsars, 
depending on the precollapse rotation, change from the 
neutrino-driven mechanism to the MHD-driven one. The magnetic effects on the 
 annihilation rates and its impacts on the dynamics are also remained to be studied
 (e.g., \citet{zhand09}).

Furthermore, the neutrino oscillation by Mikheyev-Smirnov-Wolfenstein (MSW) effect 
 (see collective references in \citet{kotake_rev,kawa09}) could be important, 
albeit in much 
later phase than we considered in this paper. It should be remembered that the 
 typical density in the polar funnel was $\rho \sim 10^{5} \gpcmc$ in the present 
model. When the density there drops as low as $\rho \lesssim 10^{3} \gpcmc$ later, 
the neutrino oscillation could operate for neutrinos traveling from the accretion
 disk to the polar funnel. If this is the 
case, the incoming neutrino spectra to the polar funnel regions and the 
pair annihilation rates there
 could be affected significantly. Together with the effects of neutrino 
self-interaction (e.g., \citet{duan06}), there seem a lot of issues 
to be clarified about 
the neutrino oscillation in the collapsar environments. As in the case 
of core-collapse supernovae 
(e.g., \citet{kotake_apjl,ott_rev} and see references therein), 
studies of gravitational-wave emissions from 
 collapsars might provide us a new window to probe into the central engine 
(e.g., \citet{hira_05,suwa_09}).
As a sequel of this work, we are planning to implement the ray-tracing calculation 
 in the hydrodynamic simulation and clarify those aspects.
 We just stand at a starting line to study those interesting 
 topics, which will be presented elsewhere one by one soon.

\acknowledgements{S.H. is grateful to T. Kajino for fruitful discussions. 
T.T. and K.K. express thanks to K. Sato, S. Yamada, and S. Nagataki
 for continuing encouragements.
Numerical computations were in part carried on XT4 and 
general common use computer system at the center for Computational Astrophysics, CfCA, the National Astronomical Observatory of Japan.  This
study was supported in part by the Grants-in-Aid for the Scientific Research 
from the Ministry of Education, Science and Culture of Japan
(Nos. S19104006, 19540309 and 20740150).

\appendix

\def\thesection{Appendix \Alph{section}}
\def\theequation{\Alph{section}\arabic{equation}}
\section{Calculations of Jacobian}
 In this section, we give an exact expression of the Jacobians 
($J_{r \mu}, J_{\mu \phi}, J_{\phi r}$) namely,
\begin{eqnarray} 
    J_{r \mu}(\mu_{\nu}, \phi_{\nu}, r_{s,\nu}, \mu_{s,\nu}) &\equiv & 
	    \left |
        \begin{array}{cc}
			\pdel{r_{s,\nu}}{\mu_{\nu} } & \pdel{\mu_{s,\nu}}{\mu_{\nu} }	\\
            \pdel{r_{s,\nu}}{\phi_{\nu}} & \pdel{\mu_{s,\nu}}{\phi_{\nu}}	\\
        \end{array}
        \right|, \label{eq:jac_start} \\
	J_{\mu \phi}(\mu_{\nu}, \phi_{\nu}, \mu_{s,\nu}, \phi_{s,\nu}) &\equiv & 
	    \left |
        \begin{array}{cc} 
			\pdel{\mu_{s,\nu}}{\mu_{\nu} } & \pdel{\phi_{s,\nu}}{\mu_{\nu} }	\\
            \pdel{\mu_{s,\nu}}{\phi_{\nu}} & \pdel{\phi_{s,\nu}}{\phi_{\nu}}	\\
        \end{array}
        \right|,  \\
    J_{\phi r}(\mu_{\nu}, \phi_{\nu}, \phi_{s,\nu}, r_{s,\nu}) &\equiv & 
	    \left |
        \begin{array}{cc}
			\pdel{\phi_{s,\nu}}{\mu_{\nu} } & \pdel{r_{s,\nu}}{\mu_{\nu} }	\\
            \pdel{\phi_{s,\nu}}{\phi_{\nu}} & \pdel{r_{s,\nu}}{\phi_{\nu}}	\\
        \end{array}
        \right|.
        \label{eq:jac0}
\end{eqnarray}
by which we can estimate the heating/momentum-transfer rates 
only by the quantities defined on the numerical grids of the hydrodynamical simulations.

From a geometrical calculation, albeit a little bit messy, the 
following partial derivatives can be found straightforward as,
 \begin{eqnarray}
	\pdel{r_{s,\nu}}{\mu_{\nu}} &=& \frac{1}{r_{s,\nu}}
			\frac{ (1 + {\mbox{\boldmath$n$}}_{s,\nu} \cdot {\mbox{\boldmath$n$}}_r)({\mbox{\boldmath$n$}}_r)_z 
			+ (|{\mbox{\boldmath$n$}}_r|^2 - 1)\cos \theta_{s,\nu} }
			{\left(1-2{\mbox{\boldmath$n$}}_{s,\nu} \cdot {\mbox{\boldmath$n$}}_r + |{\mbox{\boldmath$n$}}_r|^2\right)^{\frac{3}{2}}}, \\
	\pdel{\mu_{s,\nu}}{\mu_{\nu}} &=& \frac{\sin \theta_{s,\nu} ( 1-2{\mbox{\boldmath$n$}}_{s,\nu} \cdot {\mbox{\boldmath$n$}}_r + |{\mbox{\boldmath$n$}}_r|^2 ) 
			- \left( \cos \theta_{s,\nu} - ({\mbox{\boldmath$n$}}_r)_z \right) 
			{\mbox{\boldmath$n$}}_r \cdot {\mbox{\boldmath$k$}}_{s,\nu} }
			{ \sin \theta_{s,\nu} \left(1-2{\mbox{\boldmath$n$}}_{s,\nu} \cdot {\mbox{\boldmath$n$}}_r + |{\mbox{\boldmath$n$}}_r|^2\right)^{\frac{3}{2}}}, \\
	\pdel{\phi_{s,\nu}}{\mu_{\nu}} &=& 
			\frac{ ( \cos \theta_{s,\nu} - ({\mbox{\boldmath$n$}}_r)_z) 
			({\mbox{\boldmath$n$}}_r)_{\bot z} \cdot {\mbox{\boldmath$l$}}_{s,\nu} }
			{\left(1-2{\mbox{\boldmath$n$}}_{s,\nu} \cdot {\mbox{\boldmath$n$}}_r + |{\mbox{\boldmath$n$}}_r|^2\right)^{\frac{3}{2}}}, \\
	\pdel{r_{s,\nu}}{\phi_{\nu}} &=& \frac{1}{r_{s,\nu}}
		\frac{ \left( |({\mbox{\boldmath$n$}}_r)_{\bot z}|^2 - {\mbox{\boldmath$n$}}_{s,\nu} \cdot ({\mbox{\boldmath$n$}}_r)_{\bot z} \right) ({\mbox{\boldmath$n$}}_{s,\nu})_x 
			+ \left( \sin^2 \theta_{s,\nu} - {\mbox{\boldmath$n$}}_{s,\nu} \cdot ({\mbox{\boldmath$n$}}_r)_{\bot z} \right) ({\mbox{\boldmath$n$}}_r)_x }
			{ \left( ({\mbox{\boldmath$n$}}_r)_y - ({\mbox{\boldmath$n$}}_{s,\nu})_y \right) 
		\left( \sin^2 \theta_{s,\nu} - 2{\mbox{\boldmath$n$}}_{s,\nu} \cdot ({\mbox{\boldmath$n$}}_r)_{\bot z} + |({\mbox{\boldmath$n$}}_r)_{\bot z}|^2 \right)}, \\
	\pdel{\mu_{s,\nu}}{\phi_{\nu}} &=& \frac{-1}{\sin \theta_{s,\nu}}
		\frac{ ({\mbox{\boldmath$n$}}_r)_{\bot z} \cdot {\mbox{\boldmath$m$}}_{s,\nu}}
		{\sin^2 \theta_{s,\nu} - 2{\mbox{\boldmath$n$}}_{s,\nu} \cdot ({\mbox{\boldmath$n$}}_r)_{\bot z} + |({\mbox{\boldmath$n$}}_r)_{\bot z}|^2 }, \\
	\pdel{\phi_{s,\nu}}{\phi_{\nu}} &=& 
		\frac{ \sin^2 \theta_{s,\nu} - {\mbox{\boldmath$n$}}_{s,\nu} \cdot ({\mbox{\boldmath$n$}}_r)_{\bot z} }
		{\sin^2 \theta_{s,\nu} - 2{\mbox{\boldmath$n$}}_{s,\nu} \cdot ({\mbox{\boldmath$n$}}_r)_{\bot z} + |({\mbox{\boldmath$n$}}_r)_{\bot z}|^2 }, 
	\label{eq:jac1}
\end{eqnarray}
 where the following valuables are dependent only on the position of the neutrino sphere
 ($\theta_{s,\nu},\phi_{s,\nu}$) and the direction of a specified direction 
($\theta, \phi$) as,
 \begin{eqnarray}
	{\mbox{\boldmath$n$}}_{s,\nu} 		&=& \frac{{\mbox{\boldmath$r$}}_{s,\nu}}{|{\mbox{\boldmath$r$}}_{s,\nu}|} 
		= (\sin \theta_{s,\nu} \cos \phi_{s,\nu}, \sin \theta_{s,\nu} \sin \phi_{s,\nu}, \cos \theta_{s,\nu}), \\
	{\mbox{\boldmath$n$}}_r 		&=& \frac{{\mbox{\boldmath$r$}}}  {|{\mbox{\boldmath$r$}}_{s,\nu}|}
		= \frac{|{\mbox{\boldmath$r$}}|}{|{\mbox{\boldmath$r$}}_{s,\nu}|}(\sin \theta \cos \phi, \sin \theta \sin \phi, \cos \theta), \\
   ({\mbox{\boldmath$n$}}_r)_{z}	&=& \frac{z}          {|{\mbox{\boldmath$r$}}_{s,\nu}|}
   		= \frac{|{\mbox{\boldmath$r$}}|}{|{\mbox{\boldmath$r$}}_{s,\nu}|} \cos \theta, \\
   ({\mbox{\boldmath$n$}}_r)_{\bot z}	&=& \frac{{\mbox{\boldmath$r$}}_{\bot z}}          {|{\mbox{\boldmath$r$}}_{s,\nu}|}
		= \frac{|{\mbox{\boldmath$r$}}|}{|{\mbox{\boldmath$r$}}_{s,\nu}|}(\cos \phi, \sin \phi, 0), \\
	{\mbox{\boldmath$k$}}_{s,\nu} &=& ( \cos \theta_{s,\nu} \cos \phi_{s,\nu}, \cos \theta_{s,\nu} \sin \phi_{s,\nu}, -\sin \theta_{s,\nu}), \\
	{\mbox{\boldmath$l$}}_{s,\nu} &=& (-\sin \theta_{s,\nu} \sin \phi_{s,\nu}, \sin \theta_{s,\nu} \cos \phi_{s,\nu},  \cos \theta_{s,\nu}), \\
	{\mbox{\boldmath$m$}}_{s,\nu} &=& (-\cos \theta_{s,\nu} \sin \phi_{s,\nu}, \cos \theta_{s,\nu} \cos \phi_{s,\nu}, -\sin \theta_{s,\nu}). 
\label{eq:jac_end}
\end{eqnarray}
Finally, it should be noted that above equations can be very simple along 
the polar axis as,
\begin{eqnarray}
	\pdel{r_{s,\nu}}{\mu_{\nu}} &=& \frac{1}{r_{s,\nu}}
			\frac{ \left( \frac{Z}{r_{s,\nu}}  - \cos \theta_{s,\nu} \right) \frac{Z}{r_{s,\nu}} \cos \theta_{s,\nu}}
			{\left(1 - 2 \frac{Z}{r_{s,\nu}} \cos \theta_{s,\nu} + \left|\frac{Z}{r_{s,\nu}}\right|^2 \right)^{\frac{3}{2}}}, \\
	\pdel{\mu_{s,\nu}}{\mu_{\nu}} &=& \frac{1 - \frac{Z}{r_{s,\nu}} \cos \theta_{s,\nu}}
			{\left(1 - 2 \frac{Z}{r_{s,\nu}} \cos \theta_{s,\nu} + \left|\frac{Z}{r_{s,\nu}}\right|^2 \right)^{\frac{3}{2}}}, \\
	\pdel{\phi_{s,\nu}}{\mu_{\nu}} &=& 0, \\
	\pdel{r_{s,\nu}}{\phi_{\nu}} &=& 0, \\
	\pdel{\mu_{s,\nu}}{\phi_{\nu}} &=& 0, \\
	\pdel{\phi_{s,\nu}}{\phi_{\nu}} &=& 1,
	\label{eq:jac2}
\end{eqnarray}
where $Z$ is the distance from the equatorial plane.

\bibliographystyle{apj}
\bibliography{ms}

\end{document}